\documentclass[]{aa}
\usepackage{natbib}         
\usepackage{graphicx}
\bibpunct{(}{)}{;}{a}{}{,} 

\newcommand{\ind}[1]{_{\mathrm{#1}}}
\newcommand{\diff}{\mathrm{d}}

\def\Kepler{\emph{Kepler}}

\def\numax{\nu\ind{max}}
\def\nmax{n\ind{max}}
\def\dnuenv{\delta\nu\ind{env}}
\def\nenv{n\ind{env}}
\def\nuc{\nu\ind{c}}
\def\Dnu{\Delta\nu}
\def\dnumoy{\langle\Delta\nu\rangle}
\def\dnumoy{\Dnu}

\def\abol{\mathcal{A}\ind{bol}}
\def\corbol{\mathcal{C}\ind{bol}}
\def\modemass{\mathcal{M}}
\def\tauosc{\tau}
\def\Teff{T\ind{eff}}
\def\xB{\Lambda\ind{B}}
\def\xAKB{\Lambda\ind{A}}

\newcommand{\fnyquist}{\nu\ind{Nyquist}}
\def\visib{V^2}
\def\visiun{\visib_1}
\def\viside{\visib_2}
\def\visitr{\visib_3}
\def\amprad{\langle A_0^2 \rangle}
\def\ampav{\langle A \rangle}
\def\visitot{\visib\ind{tot}}

\def\ng{n\ind{g}}
\def\LC{\delta t\ind{LC}}

\def\Bmax{{B}_{\numax}}
\def\Hmax{{H}_{\numax}}
\def\expB{{\alpha_{B}}}
\def\smoo{{\mathcal{S}}}
\begin{document}
\title{Characterization of the power excess of solar-like oscillations in red giants with Kepler}
\titlerunning{Power excess of solar-like oscillations in red giants}
\author{B. Mosser\inst{1}\and
Y. Elsworth\inst{2}\and
S. Hekker\inst{3,2}\and
D. Huber\inst{4}\and
T. Kallinger\inst{5,6}\and
S. Mathur\inst{7}\and
K. Belkacem\inst{8,1} \and
M.J. Goupil\inst{1} \and
R. Samadi\inst{1}\and
C. Barban\inst{1}\and
T.R. Bedding\inst{4}\and
W.J. Chaplin\inst{2}\and
R.A. Garc{\'\i}a\inst{9}\and
D. Stello\inst{4}\and
J. De Ridder\inst{6}\and
C.K. Middour\inst{10}\and
R.L. Morris\inst{11}\and
E.V Quintana\inst{11}
}
\offprints{B. Mosser}

\institute{LESIA, CNRS, Universit\'e Pierre et Marie Curie, Universit\'e Denis Diderot,
Observatoire de Paris, 92195 Meudon cedex, France; \email{benoit.mosser@obspm.fr}
\and School of Physics and Astronomy, University of Birmingham, Edgbaston, Birmingham B15 2TT, United Kingdom
\and Astronomical Institute `Anton Pannekoek', University of Amsterdam, Science Park 904,
1098 XH Amsterdam, The Netherlands
\and Sydney Institute for Astronomy, School of Physics, University of Sydney, NSW 2006, Australia
\and Institute for Astronomy (IfA), University of Vienna, T\"urkenschanzstrasse 17, 1180 Vienna, Austria
\and Instituut voor Sterrenkunde, K. U. Leuven, Celestijnenlaan 200D, 3001 Leuven, Belgium
\and High Altitude Observatory, NCAR, P.O. Box 3000, Boulder, CO 80307, USA
\and Institut d'Astrophysique Spatiale, UMR 8617, Universit\'e Paris XI, B\^atiment 121, 91405 Orsay Cedex, France
\and Laboratoire AIM, CEA/DSM – CNRS - Universit\'e Paris Diderot – IRFU/SAp, 91191 Gif-sur-Yvette Cedex, France
\and Orbital Sciences Corporation/NASA Ames Research Center, Moffett Field, CA 94035, USA
\and SETI Institute/NASA Ames Research Center, Moffett Field, CA 94035, USA
}

\abstract{The space mission \Kepler\ provides us with long and
uninterrupted photometric time series of red giants. This allows
us to examine their seismic global properties and to compare these
with theoretical predictions. }
{We aim to describe the oscillation power excess observed in red
giant oscillation spectra with global seismic parameters, and to
investigate empirical scaling relations governing these
parameters. From these scalings relations, we derive new physical
properties of red giant oscillations.}
{Various different methods were compared in order to  validate the
processes and to derive reliable output values. For consistency, a
single method was then used to determine scaling relations for the
relevant global asteroseismic parameters: mean mode height, mean
height of the background signal superimposed on the oscillation
power excess, width of the power excess, bolometric amplitude of
the radial modes and visibility of non-radial modes. A method for
deriving oscillation amplitudes is proposed, which relies on the
complete identification of the red giant oscillation spectrum.
}%
{The comparison of the different methods has shown the important
role of the way the background is modelled. The convergence
reached by the collaborative work enables us to derive significant
results concerning the oscillation power excess. We obtain several
scaling relations, and identify the influence of the stellar mass
and the evolutionary status. The effect of helium burning on the
red giant interior structure is confirmed: it yields a strong
mass-radius relation for clump stars. We find that none of the
amplitude scaling relations motivated by physical considerations
predict the observed mode amplitudes of red giant stars. In
parallel, the degree-dependent mode visibility exhibits important
variations. Both effects seem related to the significant influence
of the high mode mass of non-radial mixed modes. A family of red
giants with very weak dipole modes is identified, and its
properties are analyzed.}
{The clear correlation between the power densities of the
background signal and of the stellar oscillation induces important
consequences to be considered for deriving a reliable theoretical
relation of the mode amplitude. As a by-product of this work, we
have verified that red giant asteroseismology delivers new
insights for stellar and Galactic physics, given the evidence for
mass loss at the tip of the red giant branch.}

\keywords{Stars: oscillations - Stars: interiors - Stars:
evolution - Stars: mass-loss - Stars: late-type - Methods: data
analysis}

\maketitle

\voffset = 1.5cm
\section{Introduction\label{introduction}}

The CNES CoRoT mission \citep{2006ESASP1306...39M} and the NASA
\Kepler\ mission \citep{2010Sci...327..977B} provide us with
thousands of high-precision photometric light curves of red giants
and have opened a new era in red giant asteroseismology
\citep{2009Natur.459..398D}. The asteroseismic analysis of these
data gives a precise view of pressure modes (p modes),
corresponding to oscillations propagating essentially in the large
stellar convective envelopes, as well as of mixed modes,
corresponding to pressure waves coupled to gravity waves
propagating in the core radiative regions. Previous work has
mainly dealt with the frequency properties of oscillations. In
this paper, we explore and quantify the oscillation energy. This
task is of prime importance for examining the link between
convection and oscillations, including aspects such as mode
excitation and damping. In parallel, examining the power density
of the background signal will provide useful information on
granulation.

The present work builds on previous studies based on CoRoT and
\Kepler\ observations, which derived first insights on the red
giant oscillation spectrum and provided empirical scaling
relations between the global seismic parameters
\citep{2009Natur.459..398D,2009A&A...506..465H,
2010A&A...509A..73C,2010ApJ...713L.176B,2010A&A...517A..22M,2010ApJ...723.1607H,
2011A&A...525L...9M}. Very relevant to this work,
\cite{2011arXiv1109.1194M} have determined the background signal
due to the different time scales of activity and granulation, for
the same set of \Kepler\ red giants considered here. Additionally,
\cite{2011Natur.471..608B} have recently shown with \Kepler\
observations that the evolutionary status of red giant can be
derived from the measurements of the gravity-mode spacing
observable in $\ell=1$ mixed modes \citep{2011Sci...332..205B}.
With CoRoT data, \cite{2011A&A...532A..86M} have shown that the
evolutionary status has also a clear effect on the oscillation
power excess.

In this work, we provide a first complete description of the
oscillation power excess and derive the bolometric amplitudes of
radial modes. As in previous work, we establish scaling relations
for global parameters as a function of frequency. Then, in order
to provide a direct observational comparison to theoretical
predictions \citep[e.g.][]{2009A&A...506...57D}, scaling relations
have to be established as a function of stellar parameters, such
as luminosity
\citep{1983SoPh...82..469C,1995A&A...293...87K,1999A&A...351..582H,2007A&A...463..297S}.
The influence of various parameters is tested, such as the
evolutionary status and the stellar mass.

Data, methods and definitions are presented in Sect.~\ref{method}
and in the Appendix. After the comparison of the different methods
used for measuring the global energetic parameters, scaling
relations are derived in Sect.~\ref{global}. The fine structure of
these relations is examined in Sect.~\ref{fine}. In
Sect.~\ref{ampli}, we propose a method for directly deriving the
amplitudes of radial and non-radial modes, using the complete
identification of the red giant oscillation spectra. This method
is also used in Sect.~\ref{bolo} to compare observed bolometric
amplitudes to available relationships, and in Sect.~\ref{visibi}
to determine mode visibilities.

\section{Data and method\label{method}}

\subsection{Time series}

The stars analyzed in this work have already been presented
\citep[][and references therein]{2011A&A...525A.131H}. Their
\Kepler\ magnitudes are in the range 8 -- 12.3, with most of them
fainter than magnitude 11. They show oscillations in the frequency
range [1, 280\,$\mu$Hz], for which \Kepler\ long-cadence data are
well suited, having an average sample time of 1765.5\,s, and so a
Nyquist frequency of 283.5\,$\mu$Hz. This work includes all
available data, providing time series of about 590 days for more
than a thousand red giants, giving a frequency resolution of about
19.7\,nHz. The time series were prepared from the raw data
\citep{2010ApJ...713L..87J} according to the method presented by
\cite{2011MNRAS.414L...6G}, which provides correction for various
instrumental artefacts. The mean duty cycle is 91.5\,\%, with data
loss mainly due to large but rare interruptions. In a few cases,
3-month long gaps occur for stars falling on the damaged CCD in
the \Kepler\ focal plane.

\begin{figure}
\includegraphics[width=8.88cm]{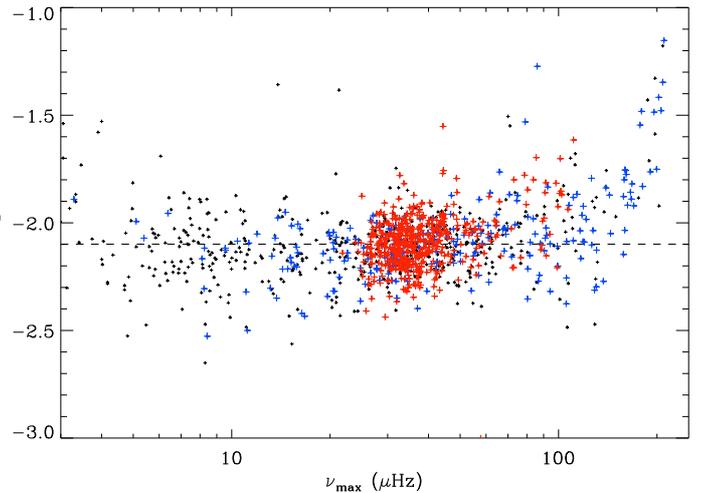}
\caption{Local value of the exponent $\expB$ of the background as
derived from Eq.~\ref{fond-local}, as a function of $\numax$. The
dashed line indicates the median value estimated in the range [5,
120\,$\mu$Hz]. The increase of $\expB$ at large frequency is an
artefact due to the Nyquist frequency. The color code indicates
the evolutionary status: clump stars in red, red giant branch
stars in blue, and undetermined status in black. \label{expofond}}
\end{figure}

\subsection{Background and power excess}

To define the different components making up the signal, we used
the classical phenomenological description of an oscillation power
excess described by a mean Gaussian profile superimposed on the
background \citep{2008Sci...322..558M}. This means that the power
density spectrum is composed of three components: the background
$B$, which is assumed to be dominated by stellar noise at low
frequency, the oscillation signal $P$ centered at $\numax$, and
white photon shot noise. The frequency $\numax$ corresponds to the
location of the maximum oscillation signal, according to the
description of the Gaussian envelope:
\begin{equation}
P (\nu) = \Hmax \exp\left[-{(\nu-\numax)^2\over 2 \sigma^2}
\right] , \label{bosse}
\end{equation}
where  $\dnuenv=2\sqrt{2\ln 2}\, \sigma$ is the full-width at
half-maximum (FWHM) of the mode envelope, and $\Hmax$ the height
at $\numax$. We note that, for the bright red giants considered
here, the photon noise component is negligible compared to the
other components.

In the global approach, the background $B$ is described by
Harvey-like components. Each component is a modified Lorentzian
\citep{1985ESASP.235..199H}:
\begin{equation}
B(\nu) = \sum_{i} {b_i \over 1 + (2\pi\,\nu\tau_i)^{\alpha_i}} ,
 \label{Harvey}
\end{equation}
where $\tau_i$ is the characteristic time scales of the $i$-th
component of maximum height $b_i$. The total number of components
varies from 1 to 3 in the different methods. Such a model provides
an accurate description of the background, although it is
phenomenological. The introduction of a Lorentzian is linked to
the assumption that the background signal is composed of
independent components with an exponential decay in the time
domain. In practice, the exponents $\alpha_i$, when fixed, are
chosen to be close to 2 or 4.

\cite{2011arXiv1109.1194M} explicitly addressed this description
of the red giant power spectra. Their work shows that the
frequency $(2\pi\tau)^{-1}$ of the granulation component close to
$\numax$ is, in fact, closely correlated with $\numax$; it varies
as $\numax^{0.89}$. In this work, we are mainly concerned with the
value $\Bmax$ of the background at $\numax$. In that respect, a
local description of $B(\nu)$ in the vicinity of $\numax$ provides
a model of sufficient precision. We have found that a polynomial
approximation of the form
\begin{equation}
B(\nu) = \Bmax \left(\nu\over \numax \right)^\expB
 \label{fond-local}
\end{equation}
is more precise than a linear fit. The expression holds in the
frequency range $[\numax-\dnuenv, \numax+\dnuenv]$. The local
parameters were estimated in the frequency ranges surrounding the
region where oscillation power excess is observed. The exponent
$\expB$ is found to be $-2.1\pm0.3$ (Fig.~\ref{expofond}).
According to the relation between $1/\tau$ and $\numax$
\citep{2011arXiv1109.1194M}, this supports an exponent close to 2
in the Harvey profile (Eq.~\ref{Harvey}), with a typical frequency
$1/\tau$ smaller than $2\ \numax$. If a modified expression
including a term in $\nu^4$ is preferred to account for a rapid
decrease of the background at higher frequency than $\numax$, the
measurement of $\expB$ close to $-2$ implies stringent conditions
between $\numax$ and $1/\tau$. We conclude that, at lower
frequencies, an exponent of 2 works reasonably well, whereas at
higher frequencies a higher power of about 4 may be required to
reproduce a steeper gradient, presumably reflecting smaller length
scales in the turbulent cascade describing the convection.

The global description of the background (Eq.~\ref{Harvey}) is
intended to be more accurate than the local description
(Eq.~\ref{fond-local}). However, until we have a physical
description of the background and its different scales, the global
description remains phenomenological. We must keep in mind that
crosstalk between the different components of the signal might add
or remove energy from the oscillation signal. In that respect,
testing the local description of the background, which is
certainly not a correct solution, offers the possibility to fit
the background with a minimum number of components.

In order to obtain the global parameters of the Gaussian envelope,
most of the methods used for this work first consider a smoothed
spectrum. The best practice for achieving this step is described
in Appendix \ref{appendix-normalisation} and
\ref{appendix-smoothing}.

\begin{figure*}
\includegraphics[width=17cm]{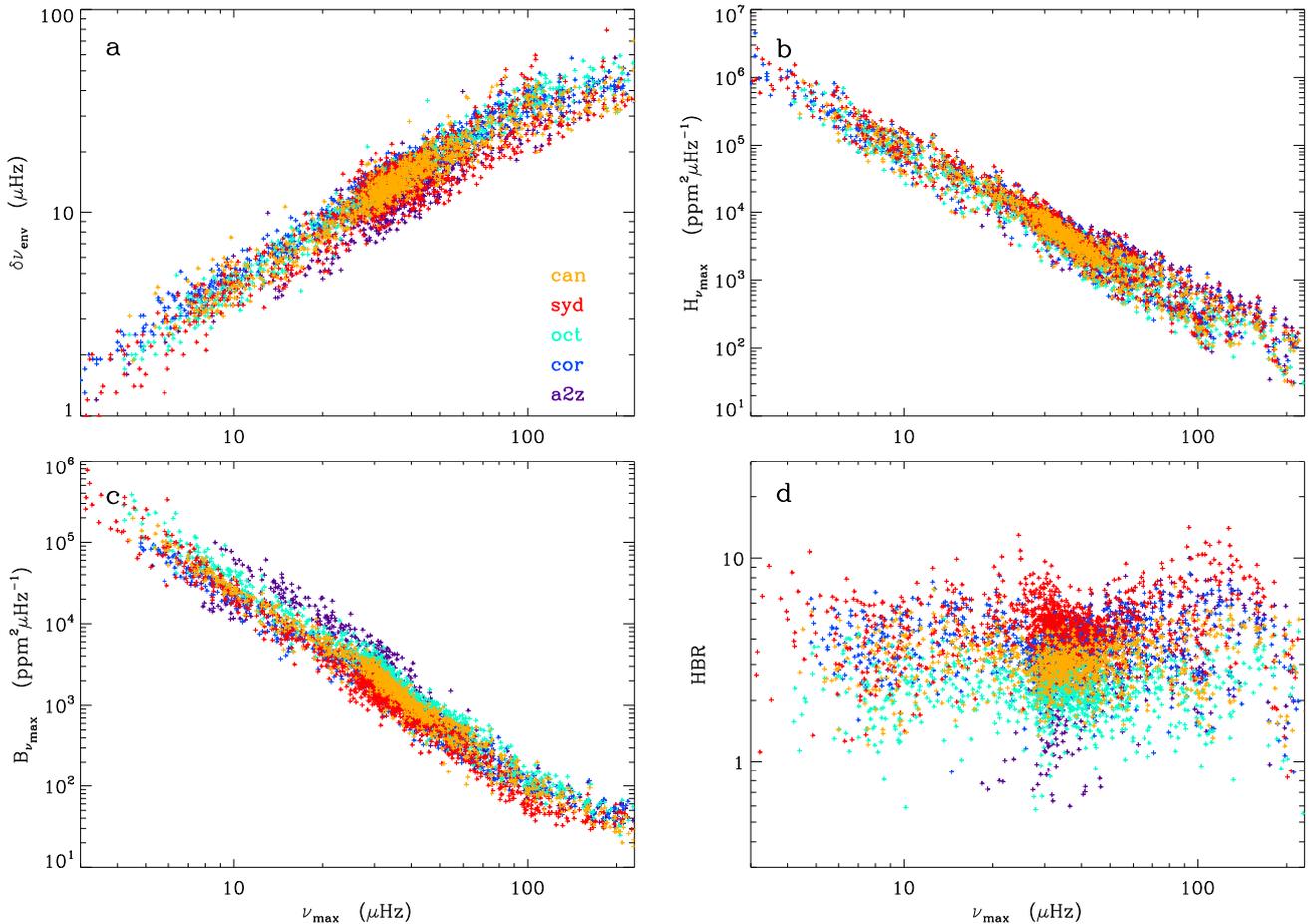}
\caption{Global seismic parameters as a function of $\numax$
derived from the five methods used here.  {\bf a)} FWHM $\dnuenv$
of the Gaussian envelope; {\bf b)} Mode height $\Hmax$, as defined
by Eq.~\ref{bosse}, corrected for apodization; {\bf c)} Background
at $\numax$. The increase above 200\,$\mu$Hz is  an artefact due
to the proximity of the Nyquist frequency. {\bf d)}
Height-to-background ratio, affected by the artefact on $\Bmax$.
The color code indicates the method used for the data analysis, as
indicated in panel {\bf a)}. The deviations from the trends above
200\,$\mu$Hz are due to the proximity to the Nyquist frequency.
\label{fig-compar}}
\end{figure*}

\section{Scaling relations for global energy parameters\label{global}}

\subsection{Comparison of the methods}

We focus on the three global parameters previously introduced for
describing the power excess (Eqs.~\ref{bosse}-\ref{fond-local}):
\begin{itemize}
    \item The FWHM $\dnuenv$ of the Gaussian envelope, which measures
    the extent of efficient excitation of the oscillations.
    As we already know from CoRoT observations, the FWHM increases
    significantly with $\numax$ (Fig.~\ref{fig-compar}a). This
    envelope width is important to estimate
    the number $\nenv = \dnuenv / \dnumoy$ of radial orders with detectable
    amplitudes. This number can be compared to the order of
    maximum signal, defined as $\nmax = \numax/ \dnumoy - \varepsilon
    (\dnumoy)$.
    \item The oscillation height $H$ is measured at $\numax$ (Fig.~\ref{fig-compar}b).
    \item The background signal underlying the oscillation signal is taken as the
    height $\Bmax$ at $\numax$ (Fig.~\ref{fig-compar}c). From
    $\Bmax$, we can derive the height-to-background  ratio $\Hmax / \Bmax$
    (Fig.~\ref{fig-compar}d).
\end{itemize}

The preliminary normalization process (Appendix
\ref{appendix-normalisation}) ensures all methods produced results
in global agreement. Small differences nevertheless remain, mostly
being related to the way the background is modelled, particularly
the number of components in the background profile. The
differences have a stronger impact on the multiplicative
coefficients than on the exponents of the scaling relations.
Despite the correction of the time-domain averaging effect towards
high frequencies due to the half-hour integration time, expressed
by the suppression factor $\hbox{sinc\,}(\pi\numax / 2
\fnyquist)$, we note that all methods are affected by the Nyquist
frequency. When $\numax$ is close to $\fnyquist$, it becomes
difficult to estimate the background, and unsurprisingly all
methods tend to underestimate it since fewer modes are observable.
As a result, the following analysis is limited to stars with
$\numax < 200\,\mu$Hz.

The parameter showing the best agreement between the different
methods is the envelope width $\dnuenv$, while the parameter with
the largest spread is the height-to-background ratio. This
reflects systematic variations in the way the different components
of the Fourier spectrum were fitted. We note systematic
differences in $\Hmax$ and $\Bmax$, often with anticorrelated
variations. We also note a change at the clump-frequency, which is
certainly real, and is studied in detail in Sect.~\ref{fine}.
Finally, we consider that the local description of the background
provides results intermediate between the different global
descriptions based on Harvey profiles. We therefore chose to use
it for analyzing the scaling relations in the following sections.

\begin{table}
\caption{Comparison of global seismic
parameters}\label{comp-methode}
\begin{tabular}{lcccc}
\hline
method  &$\dnumoy$&$\dnuenv$&$\Hmax/10^7$&$\Bmax/10^6$\\
\hline
    A2Z & 0.269;0.755 & 0.49;0.89 & 0.90;$-$2.10 & 11.1;$-$2.56 \\
    CAN & 0.292;0.740 & 0.71;0.82 & 1.95;$-$2.35 & 5.75;$-$2.31 \\
    COR & 0.286;0.744 & 0.74;0.84 & 1.69;$-$2.28 & 3.76;$-$2.23 \\
    OCT & 0.271;0.756 & 0.57;0.90 & 1.74;$-$2.38 & 8.16;$-$2.37 \\
    SYD & 0.283;0.747 & 0.63;0.85 & 1.95;$-$2.32 & 5.36;$-$2.39 \\
\hline
\end{tabular}

- Each column gives a doublet $\alpha ; \beta$, respectively
coefficient and exponent of the power laws $\alpha\, \numax^\beta$
of each parameter.

- Frequencies are all expressed in $\mu$Hz, and heights in
ppm$^2\mu$Hz$^{-1}$.

- The different methods provided results for a number of targets
varying between 720 and 1220.

\end{table}

\subsection{Scaling relations}

Scaling relations for $\dnuenv$, $\Hmax$ and $\Bmax$ as a function
of $\numax$ show clear global trends (Table~\ref{comp-methode},
Fig.~\ref{fig-compar}). Mean values of the coefficients and
exponents of the best fits derived with all methods are given in
Table~\ref{table-fit}. The dispersion with respect to the power
laws, even if larger than for the $\numax - \dnumoy$ relation, are
fairly small. Many reasons, with different origins, can explain
this. Only physical reasons related to the seismic properties are
discussed below. Other factors, such as the influence of
unidentified close companions or background objects on the red
giant light curves, are not relevant in this study since the red
giants currently accessible were chosen to be bright and in
uncrowded fields.

\begin{itemize}
\item The scaling relation for $\dnuenv$ observed for CoRoT
targets is confirmed. The `saturation' above 200\,$\mu$Hz is
certainly related to the difficulty in estimating the background
when $\numax$ is close to $\fnyquist$. As stated earlier, stars
with $\numax$ greater than 200\,$\mu$Hz were not considered in the
following scaling relations. But we also note a saturation in the
frequency range [100, 200\,$\mu$Hz] that cannot be explained by
this artefact. This effect is studied in detail in
Sect.~\ref{fine}. \item The number of observable degrees varies
    as $\dnuenv / \dnumoy$. It decreases
    slowly at low frequency since it scales as $\numax^{0.13}$ \citep{2010A&A...517A..22M}.
\item The stellar background signal varies as $\numax^{-2.4}$. The
exponent differs from the value $-2$ observed for subgiants and
    main-sequence stars \citep{2011ApJ...732L...5C}.
\item The exponents of the oscillation and background heights are
approximately equal. As a consequence, the HBR appears to be
constant with frequency (Fig.~\ref{fig-compar}d). This was not the
case in CoRoT data, where the background for many stars was
dominated by the stellar photon noise, without a possibility to
disentangle precisely the photon noise component from the stellar
background. Here, photon noise is negligible compared to the
stellar background for the majority of the targets. The HBR at
$\numax$ measures the energy density of the oscillations relative
to the stellar background. The constant ratio, equal to about 3.8,
is discussed in the next section. We note that the HBR values in
Fig.~\ref{fig-compar}d have a high dispersion in the frequency
range [30, 60 \,$\mu$Hz]. The very large number of objects
observed in this range excludes spurious variation due to a poor
sampling: these stars definitely present characteristics that
cannot be described by a single power law in $\numax$.
\end{itemize}
For the remainder of the discussion, all results are based, for
consistency, on a single method (COR, Table \ref{pipelines}).

\begin{table}
\caption{Scaling relations}\label{table-fit}
\begin{tabular}{lccc }
\hline parameter  & coefficient $\alpha$ & exponent $\beta$\\
\hline
$\dnumoy$  &$0.276\pm 0.002$      & $0.751\pm0.002$\\
$\nmax=\numax/\Dnu$&$3.52\pm0.004$      & $0.253\pm0.002$\\
$\dnuenv$  &$0.66 \pm 0.01 $      & $0.88\pm0.01$  \\
$\nenv= \dnuenv/\Dnu$&$2.33\pm 0.01$     & $0.13\pm0.01$  \\
$\Hmax$    &$(2.03\pm 0.05)\;10^7$& $-2.38\pm0.01$ \\
$\Bmax$    &$(6.37\pm 0.02)\;10^6$& $-2.41\pm0.01$ \\
\hline
\end{tabular}

- Mean values from the combined results of all methods.

- Each parameter is estimated as a power law of $\numax$, namely
varying as $\alpha\, \numax^\beta$, with $\alpha$ the coefficient
and $\beta$ the exponent.

- Frequencies are all expressed in $\mu$Hz, and heights in
ppm$^2/\mu$Hz.

- Scaling relations concerning $\dnuenv$ are only valid for stars
with $\numax$ less than 100\,$\mu$Hz.
\end{table}

\section{Fine detail\label{fine}}

\begin{figure}
\includegraphics[width=8.88cm]{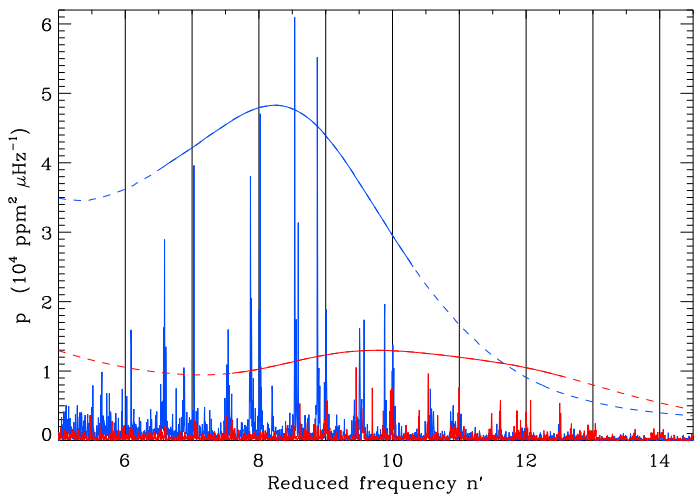}
\caption{Comparison of two stars with similar large separations,
one belonging to the RGB (KIC 4750456, $\dnumoy = 5.89\,\mu$Hz,
blue curve) and the other to the secondary clump (KIC 3758458,
$\dnumoy = 5.87\,\mu$Hz, red curve). The x-axis is the reduced
frequency $n'$; vertical lines indicate radial modes. The dashed
curves represent 20 times the smoothed spectra; the regions
corresponding to the FWHM are overplotted with solid lines.
\label{compar_amp}}
\end{figure}

\begin{figure}
\includegraphics[width=8.88cm]{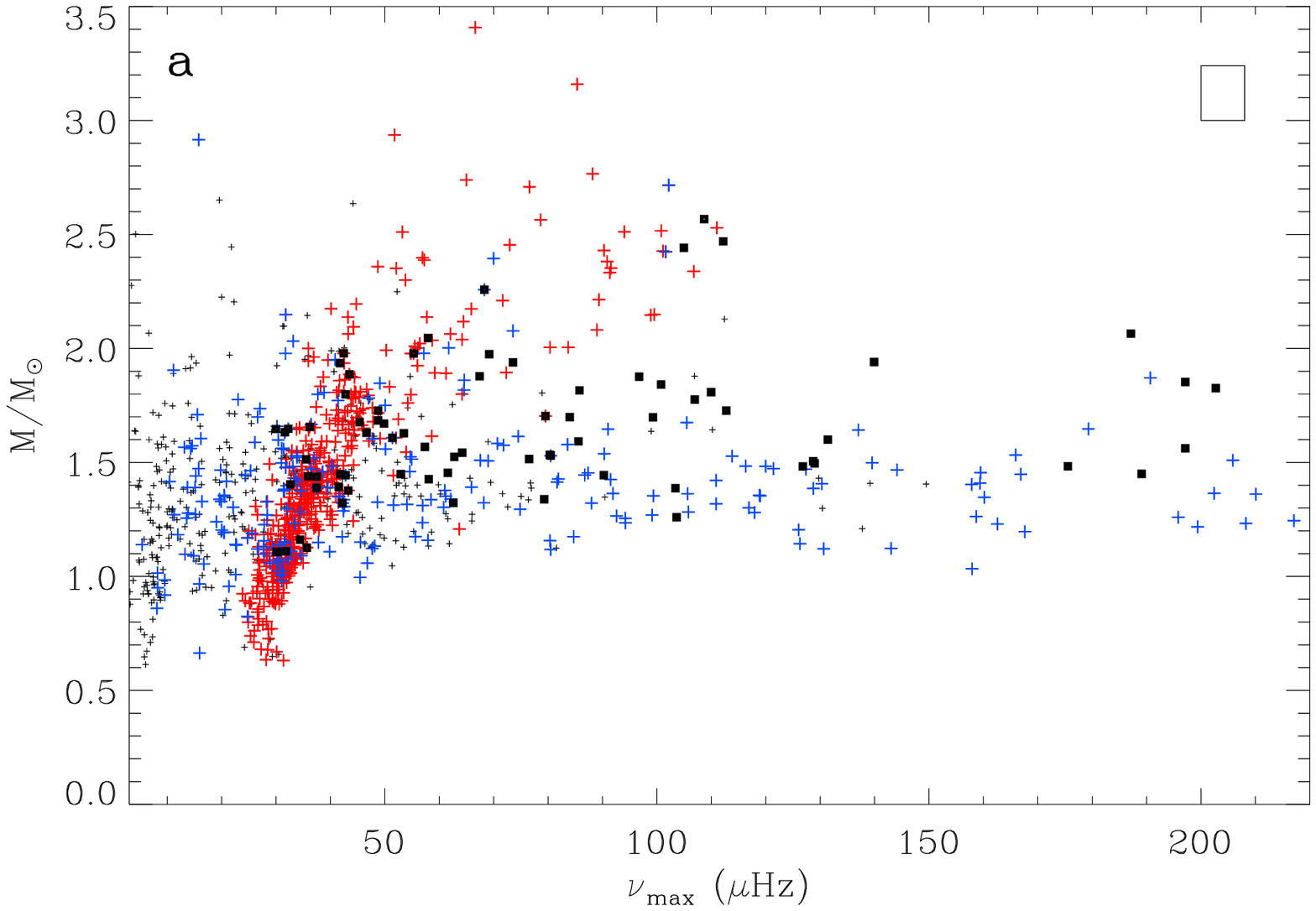}
\includegraphics[width=8.88cm]{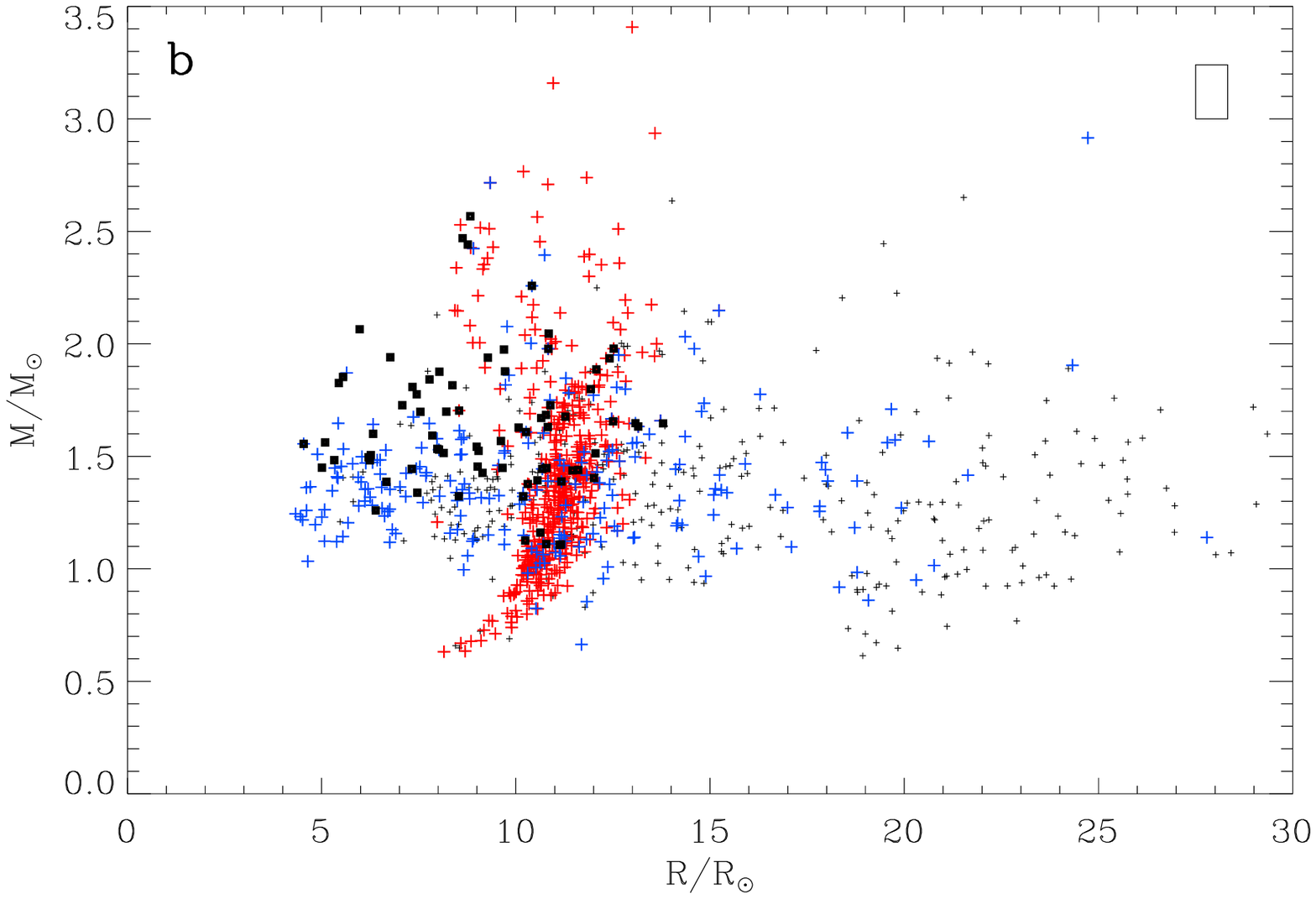}
\caption{Asteroseismic mass as a function of $\numax$ (panel {\bf
a}) and of the asteroseismic radius (panel {\bf b}). The color
code indicates the evolutionary status; clump stars in red, giant
branch stars in blue, unknown status in dark grey. The population
of giants with low $\ell=1$ amplitude (identified later in
Sect.~\ref{veryweak}) is indicated with black squares. The
rectangles in the upper right corners indicate the mean value of
the 1-$\sigma$ error bars. \label{correl-masse}}
\end{figure}

\begin{figure}
\includegraphics[width=8.88cm]{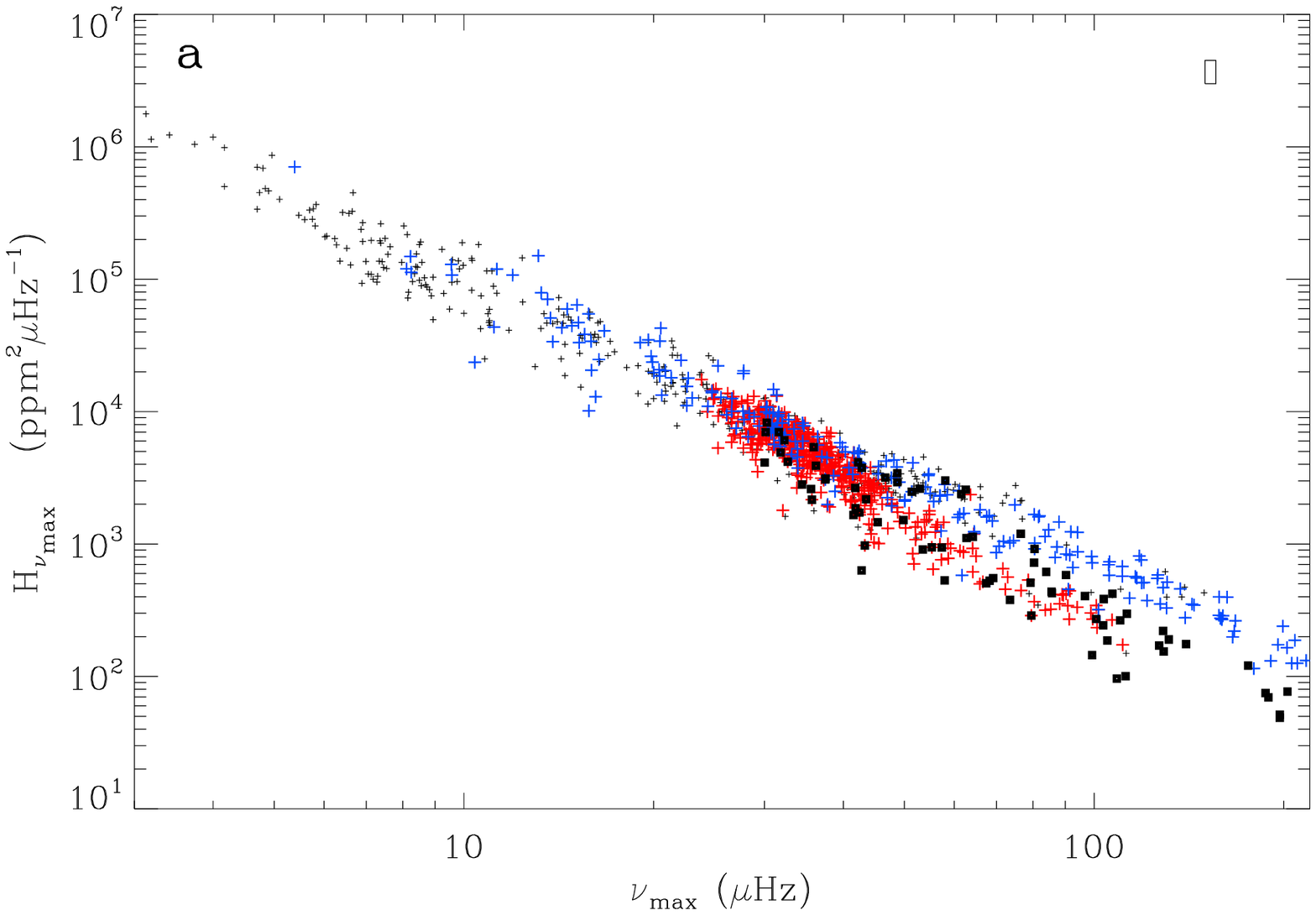}
\includegraphics[width=8.88cm]{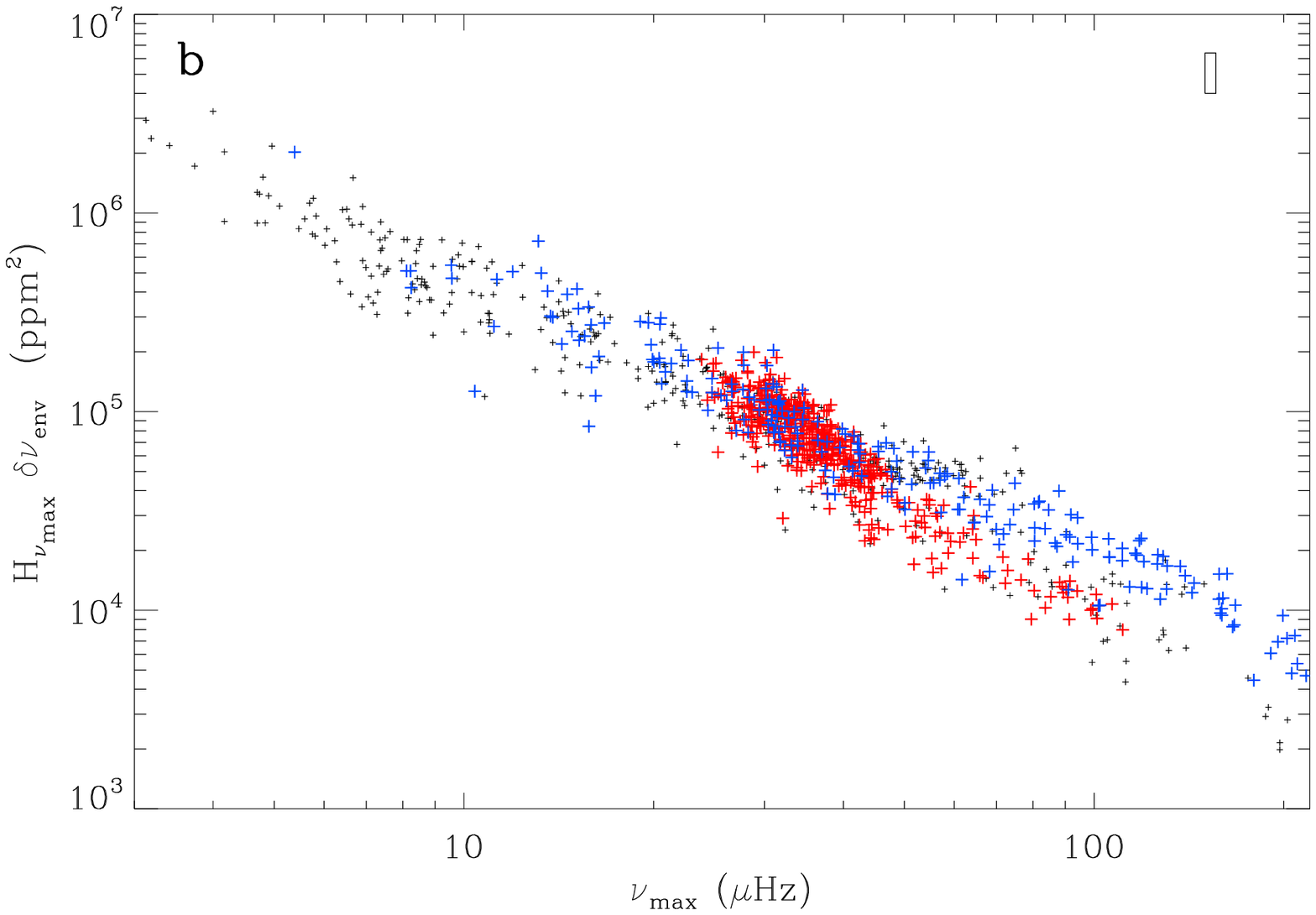}
\caption{Maximum height $\Hmax$ at $\numax$ (panel {\bf a}) and
product $\Hmax \dnuenv$ (panel {\bf b}), as a function of
$\numax$. Evolutionary status and mean error bars are indicated as
in Fig.~\ref{correl-masse}. \label{height_evolution}}
\end{figure}

\begin{figure}
\includegraphics[width=8.88cm]{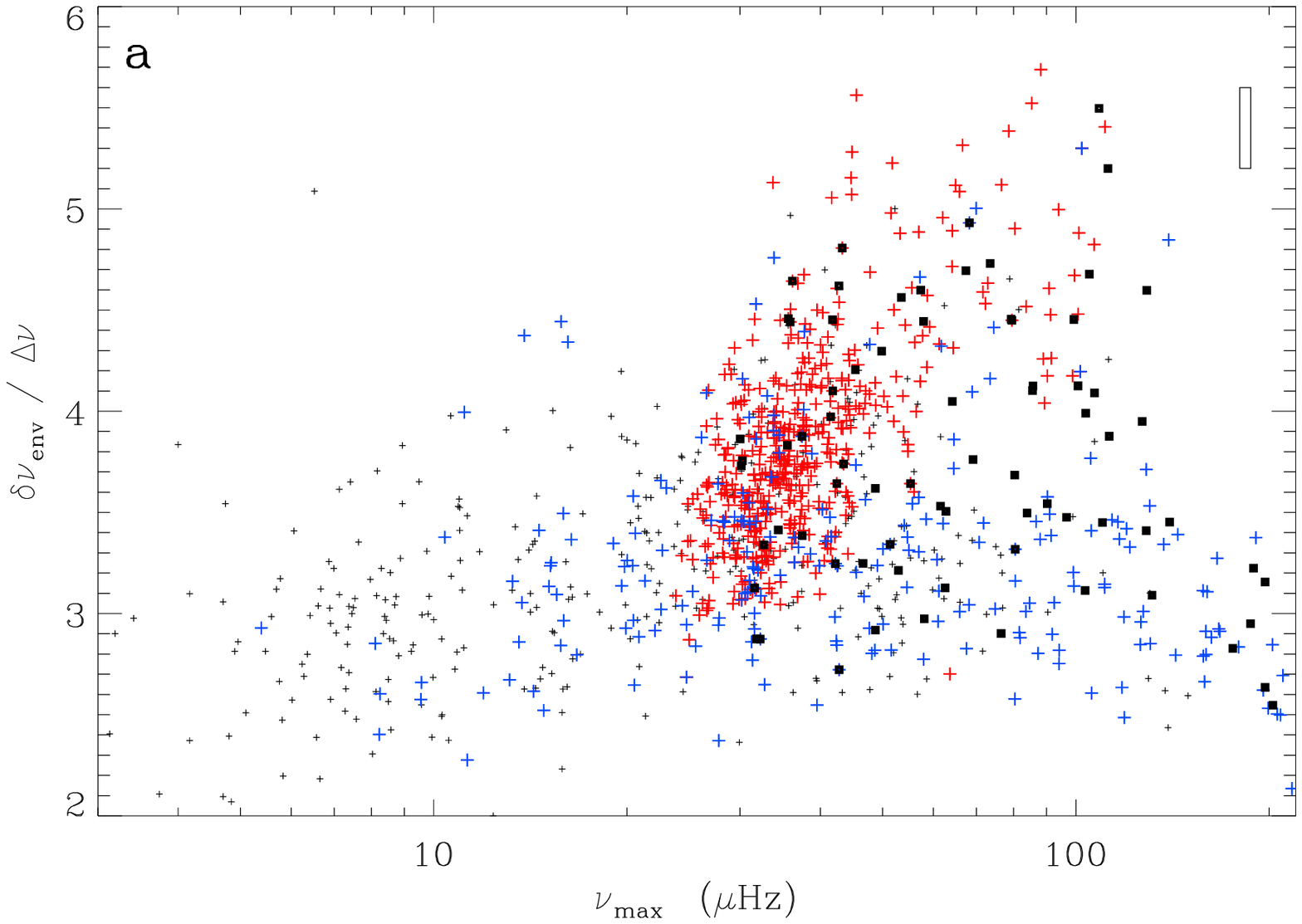}
\includegraphics[width=8.88cm]{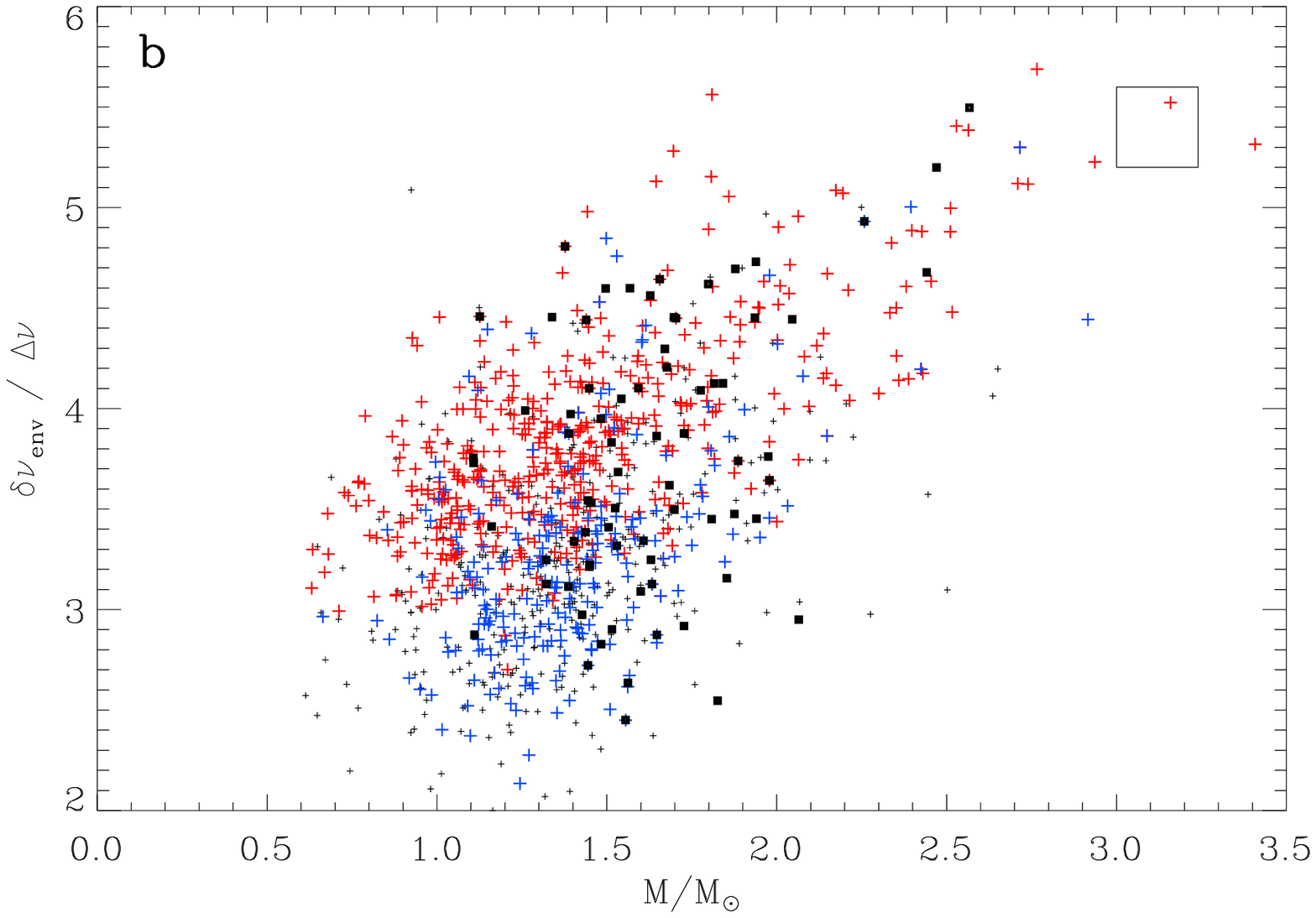}
\caption{FWHM in unit $\Dnu$ as a function of $\numax$ (panel {\bf
a}) and of the seismic mass (panel {\bf b}). Evolutionary status
and mean error bars are indicated as in Fig.~\ref{correl-masse}.
\label{env_evolution}}
\end{figure}

Since we lack theoretical explanations for the observed scaling
relations, we examine in this section how these relations are
influenced by stellar parameters. The role of the stellar mass is
potentially important since the mass distribution is degenerate
with $\numax$ \citep{2010A&A...517A..22M}. So, we have derived an
estimate of the asteroseismic mass from the scaling relations for
$\dnumoy$ and $\numax$ to address this mass influence.

In a next step, we have examined whether the relations discussed
in Sect.~\ref{global} vary with the evolutionary status. The
direct comparison of two stars with similar large separations but
different evolutionary status, one belonging to the red giant
branch and the other to the red clump, clearly shows that the
clump stars present much lower amplitudes, but larger values of
$\numax$ and $\dnuenv$, hence a higher acoustic cutoff frequency
(Fig.~\ref{compar_amp}). The relation between $\numax$ and the
cutoff frequency has been shown observationally, and is now
assessed theoretically \citep{2011A&A...530A.142B}.

\subsection{Mass-radius relation}

In order to investigate how global seismic parameters vary with
both mass and evolutionary status, we must illustrate how the
mass, radius and evolutionary status are linked.
Fig.~\ref{correl-masse} shows how the stellar masses and radii are
distributed with $\numax$ and how they are correlated with the
evolutionary status. For stars on the red giant branch (RGB), the
mass distribution is nearly uniform across the whole frequency
range. On the other hand, the distribution of $\numax$ and mass
are clearly correlated for clump stars, as already shown with
CoRoT data. The same features are clearly found in the mass-radius
relation, since stellar radius and $\numax$ are strongly
correlated. A discussion in terms of stellar evolution, mass loss,
and reset of the red giant structure at the helium flash for
low-mass stars, has already been presented by
\cite{2011A&A...532A..86M}. Here, the clear difference between RGB
and clump stars helps to identify the structure present in
Fig.~\ref{fig-compar}b: the apparent slope change in the $\Hmax$
and $\Bmax$ relations is linked to the highly non-uniform mass
distribution for $\numax$ in the frequency range [30, 60\,$\mu$Hz]
due to clump stars.

\subsection{Power excess, mass, and evolutionary status\label{statut}}

We have revisited the scaling relations dealing with the power
excess according to the evolutionary status of the red giants
(Figs.~\ref{height_evolution} and \ref{env_evolution}). This
reveals a clear difference between the two populations.

\begin{itemize}
    \item For stars with $\numax$ in the range [25-50\,$\mu$Hz]: red clump
stars and RGB stars have similar masses, but the distribution of
the mass on the RGB is not correlated with $\numax$, whereas there
is a clear correlation for red-clump stars
(Fig.~\ref{correl-masse}). At the same time, the global energy
parameters are similar for both populations, but with somewhat
different slopes (Fig.~\ref{height_evolution}).
    \item At higher $\numax$ we find
    the secondary clump, consisting of He-burning stars of higher masses
    \citep{1999MNRAS.308..818G}. With increasing
    mass, the maximum mode height of secondary clump stars
    decreases significantly. In parallel, the FWHM of
    the Gaussian envelope increases. In stars burning helium in their core,
    the energy partition differs
    significantly: a wider range of modes is excited
    but with much lower oscillation heights.
    This can be understood when recalling the link between $\numax$
and the cutoff frequency $\nuc$
\citep{1995A&A...293...87K,2011A&A...530A.142B}. At fixed
$\dnumoy$ (as for the two red giants shown in
Fig.~\ref{compar_amp}), a low-mass star has a much lower $\numax$
than a member of the secondary clump. The explicit mass dependence
of the FWHM $\dnuenv$ is shown in Fig.~\ref{env_evolution}. The
    product ${\Hmax \dnuenv}$ (Fig.~\ref{height_evolution}b) tells us that oscillations in
    clump stars have less energy than in the RGB.
\item
    The saturation of $\dnuenv$ observed in the frequency
     range [100 - 200 $\mu$Hz] only concerns RGB stars. It occurs at too low a frequency
     to be related to the Nyquist frequency, and clearly
     departs from the scaling relation reported in Table~\ref{comp-methode}.
     We have to conclude that the mechanism of excitation is only
     efficient in a limited frequency range. This may be due to a
     change in the relation between $\numax$ and the
     atmospheric cutoff frequency \citep{2011A&A...530A.142B}.
     Finally, we can derive a relation between $\dnuenv$ and the stellar
     mass, expressed by the number of observable radial orders:
\end{itemize}
\begin{equation}\label{rapport_nuenv}
 \nenv \equiv {\dnuenv \over \Dnu} =
 \left\{\begin{array}{rrl}
    2.72 &+\ 0.87\ M/M_\odot & \qquad\hbox{(clump)} \\
    2.37 &+\ 0.77\ M/M_\odot & \qquad\hbox{(RGB)} \\
 \end{array}\right.
\end{equation}

\begin{figure}
\includegraphics[width=8.88cm]{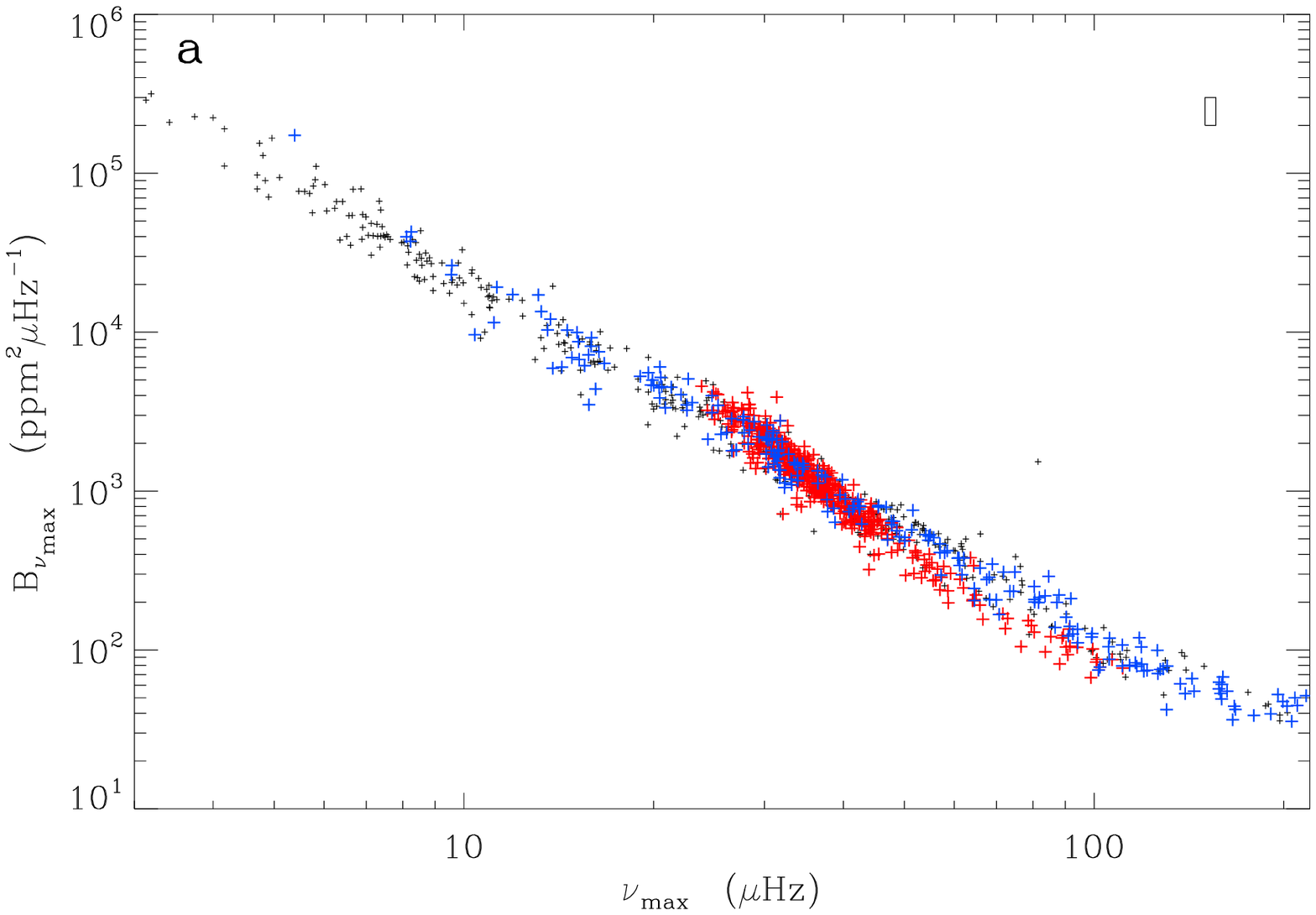}
\includegraphics[width=8.88cm]{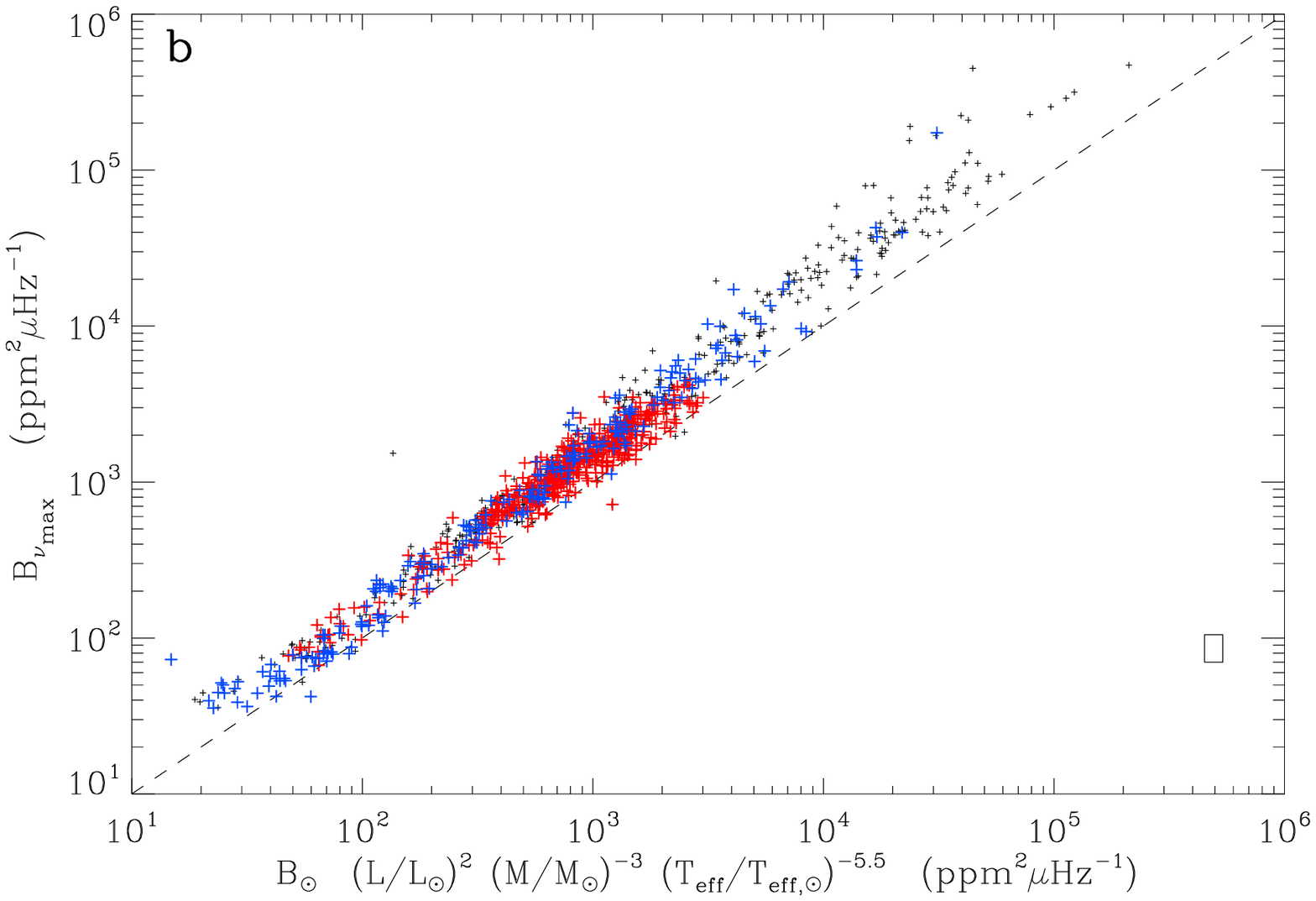}
\caption{Background $\Bmax$ as a function of $\numax$ (panel {\bf
a}), and as a function of $B_\odot \, \xB$ (panel {\bf b}), with
the solar value $B_\odot = 0.19$\,ppm$^2\,\mu$Hz$^{-1}$.
Evolutionary status and error bars are indicated as in
Fig.~\ref{correl-masse}. The dashed line indicates the 1:1
relation. \label{Bmax_evolution}}
\end{figure}

\subsection{Granulation background}

We have tested the new relation giving $\Bmax$ as a function of
$\xB = (L/L_\odot)^2  (M/M_\odot)^{-3} (\Teff/T_\odot)^{-5.5}$
\citep{2011A&A...529L...8K}. We made use of the asteroseismic
estimate of the luminosity derived from the asteroseismic radius,
the effective temperature \citep{2011AJ....142..112B} and the
Boltzmann law. Such luminosity values certainly suffer from
scaling uncertainties. This means that the absolute values are
less reliable than relative variations. In other words, the
exponent of the relation predicting $\Bmax$ is less biased than
the multiplicative coefficient.

The function $\Bmax (\numax)$ shows two different branches with
different slopes (Fig.~\ref{Bmax_evolution}a), whereas $\Bmax
(\xB)$ seems to be able to reconcile both evolutionary states
(Fig.~\ref{Bmax_evolution}b). The phenomenological expectation
seems verified, since the exponent is not far from unity, in
agreement with \cite{2011arXiv1109.3460H}:
\begin{equation}\label{rapportHB}
 \Bmax =
 \left\{\begin{array}{rll}
   0.16 & \xB^{1.08} &\qquad\hbox{(all)} \\
   0.28 & \xB^{1.01} &\qquad\hbox{(clump)} \\
   0.15 & \xB^{1.09} &\qquad\hbox{(RGB)} \\
 \end{array}\right.
\end{equation}
We note that the coefficient of the fit of all red giants is very
close to the value of the solar background at $\numax$ of about
0.19\,ppm$^2 \, \mu$Hz$^{-1}$ found by \cite{2011arXiv1109.3460H}.
This value has been used in Fig.~\ref{Bmax_evolution}b. We see
that the relation predicts too low values of the background at
$\numax$.

Having previously remarked that the HBR has a uniform value across
a wide frequency range, we have then to examine whether $\Hmax$
also scales with $\xB $. This is the case, with exponents close to
1 for both RGB and clump stars, but with a different ratio. We
measure:
\begin{equation}\label{rapportHB}
 {\Hmax \over \Bmax} =
 \left\{\begin{array}{cl}
   3.67 \pm 0.07 & \qquad\hbox{(clump)} \\
   4.02 \pm 0.12 & \qquad\hbox{(RGB)} \\
 \end{array}\right.
\end{equation}
\cite{2011A&A...529L...8K} postulated that the mean height of
modes, observed in velocity, is proportional to the velocity power
density of the granulation at $\numax$. Here, we show that this
proportionality is verified in the photometric signal for red
giants. A similar study for all stellar classes was made by
\cite{2011arXiv1109.3460H}, who reached similar conclusions.

\section{Amplitudes\label{ampli}}

It has been suggested that oscillation amplitudes depend on the
luminosity-to-mass ratio
\citep{1983SoPh...82..469C,1995A&A...293...87K,1999A&A...351..582H,2007A&A...463..297S}.
They also depend on the mode degree. As we found a large diversity
in the red giant oscillation spectra, we propose in this section a
method for measuring separately the radial and non-radial
amplitudes in red giants facilited by the automated mode
identification provided by the red giant universal oscillation
pattern (Appendix \ref{appendix-identification}).
The measurements will then allow us to derive bolometric
amplitudes and also mode visibilities, in Sect.~\ref{bolo} and
\ref{visibi} respectively.

\begin{figure*}
\includegraphics[width=17cm]{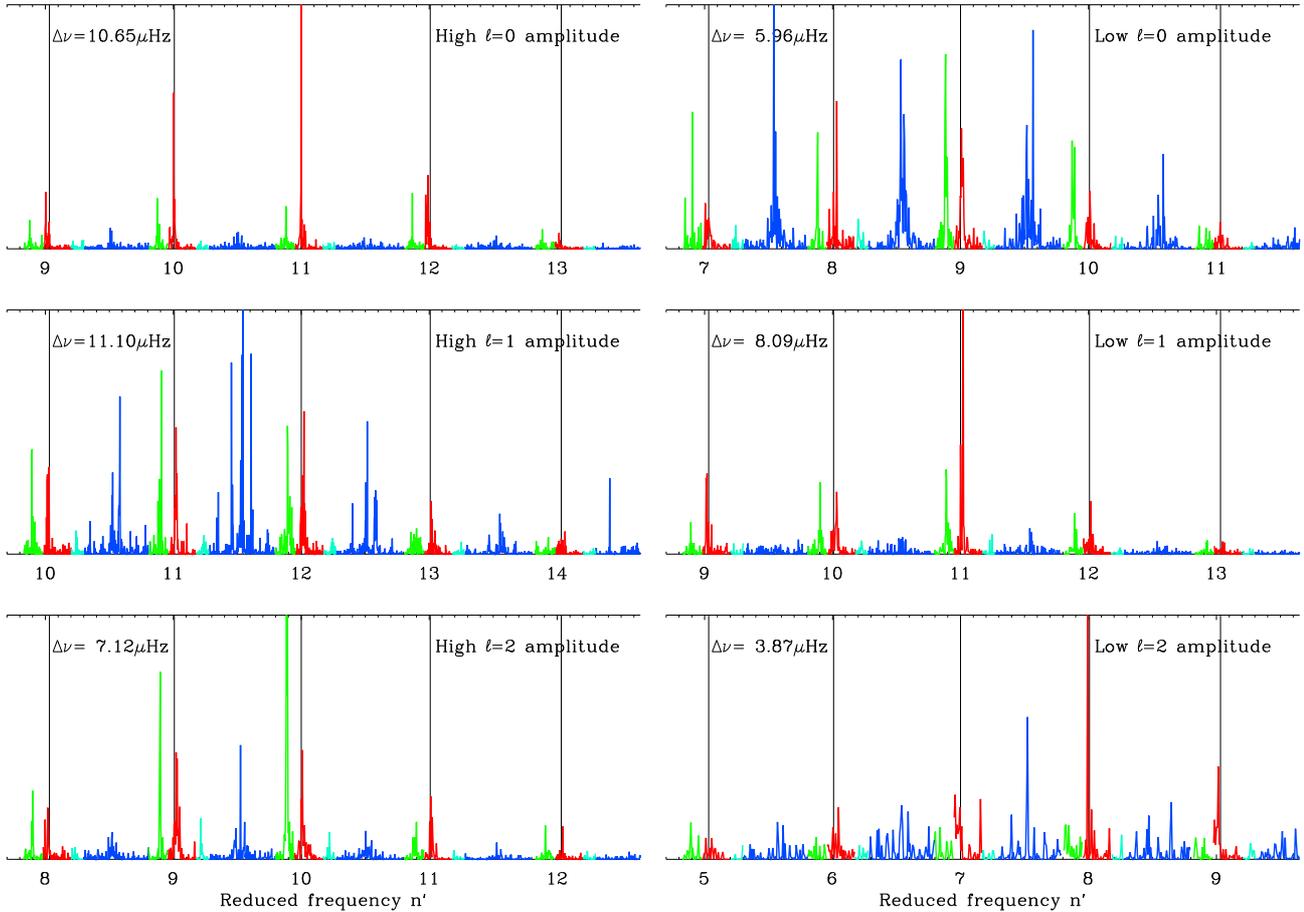}
\caption{Identification of the ridges in red giants observed by
\Kepler. Targets were chosen with oscillation spectra exhibiting
very different visibility coefficients (see Sect.~\ref{visibi}).
Left side: high visibility of $\ell$=0 to 2 from top to bottom.
Right side: low visibility of $\ell$=0 to 2 from top to bottom.
The spectra are corrected for the background contribution
(negative values being omitted), and plotted as a function of the
reduced frequency. The color code, derived from Eqs. \ref{calculA}
and \ref{calculAl}, presents radial modes in red, $\ell=1$ in
blue, $\ell=2$ in green, and $\ell=3$ in light blue.
\label{difference}}
\end{figure*}

\subsection{Radial amplitudes}

Previously, the amplitudes of radial modes have been derived from
the mean amplitude in the oscillation envelope divided by the
total visibility factors of the modes
\citep{2008ApJ...682.1370K,2009A&A...495..979M}. Here, we have
computed spectra as a function of the reduced frequency $n'$
(Eq.~\ref{reduit}). Squared amplitudes are then given by
integrating the height across the mode width, as estimated from
the fit of the ridges, with the integral being corrected for the
background contribution. That is, we have computed the squared
amplitudes for each radial order from:
\begin{equation}\label{calculA}
    A_0^2(n) = \delta \nu\ \int_{n-e_{20}}^{n+e_{03}} \bigl[ p(n') - B  \bigr]\ \diff n'\
\end{equation}
with $p$ the power density spectrum, $\delta\nu$ the frequency
resolution, and $B$ the local background. The boundaries $e_{20}$
and $e_{03}$ bracket the frequency range where radial modes are
expected: the radial mode $(n,0)$ lies between the $(n-1,2)$ and
$(n-1,3)$ modes. They have been defined according to the
parametrization proposed by \cite{2011A&A...525L...9M}. We have
checked that the measures of the radial amplitudes are stable with
respect to slight modifications of these boundaries. The major
contribution to the uncertainties comes from uncertainty on the
background correction, resulting in a mean precision of about
10\,\%.

\subsection{Non-radial amplitudes}

Non-radial amplitudes have been calculated in the same way, with
the appropriate boundaries $e_{\ell\ell'}$ between the adjacent
degrees $\ell$ and $\ell'$:
\begin{equation}\label{calculAl}
    A_\ell^2(n) = \delta \nu\ \int_{n-e_{\ell'\ell}}^{n+e_{\ell\ell''}}
    \bigl[ p(n') - B  \bigr]\ \diff n'\ .
\end{equation}
We have used the mean values  $e_{12} = -0.22$, $e_{20} = -0.065$,
$e_{03} = 0.17$, and $e_{31} = 0.27$ derived from the universal
red giant oscillation pattern \citep{2011A&A...525L...9M}. These
boundaries are globally shifted by $0.008\, (n-\nmax)$ to account
for the mean curvature of the spectra, and modulated in large
separation according to the exact location of the ridges
\citep[Table 1 of][]{2011A&A...525L...9M}. In case of mixed modes,
the amplitude of a certain degree we measure corresponds to the
sum of all individual components of a given pressure radial order
$n$. A few examples are given in Fig.~\ref{difference}.

From the definition given by Eq.~\ref{calculAl}, it is clear that,
in case of rotational multiplets, the method integrates the
squared amplitudes of all components. In other words, the method
is not affected by the unknown stellar inclination.

\subsection{Mean values of individual amplitudes}

The mean value of the individual amplitudes has been obtained
under the assumption that the energy partition is Gaussian,
following Eq.~\ref{bosse}:
\begin{equation}\label{calculAmoy}
    \langle A_\ell^2 \rangle = \displaystyle{\sum_{\nmax-2}^{\nmax+2} A_\ell^2(n)}
    \ \Big{/}\
\displaystyle{\sum_{\nmax-2}^{\nmax+2}
\exp\left[-{(\nu_{n,\ell}-\numax)^2\over 2 \sigma^2} \right]}
 . \label{boss}
\end{equation}
Formally, one would expect the weighting to be performed before
the summation, but such a method is too sensitive to the varying
amplitudes created by the stochastic excitation. Therefore, with
Eq.~\ref{calculAmoy}, we intend to estimate a mean value not
affected by this source of noise. Considering five radial orders
is in agreement with the observed width of the envelope and allows
us to pick the highest peaks.

\subsection{Individual and globally averaged radial amplitudes}

The radial amplitude has also been directly determined, by
\begin{equation}\label{}
    A_0 = A_0(\nmax).
\end{equation}
We have compared $\amprad^{1/2}$ to the radial amplitude $A_0$,
and shown that both amplitudes are closely linked, with $A_0
\simeq 1.07 \ \amprad^{1/2}$. However, to avoid scattering due to
the finite mode lifetimes and for coherence of the visibility
calculation, we have decided to consider the radial amplitude
$\amprad^{1/2}$. This value can be compared to the non-radial
amplitudes in order to compute the mode visibilities. It can also
be compared to the global average amplitudes derived from the
total mode visibility
\citep{2008ApJ...682.1370K,2009A&A...495..979M}
\begin{equation}\label{average}
    \ampav = \sqrt{\Hmax \Dnu \over \visitot} .
\end{equation}
The examination of the total mode visibility $\visitot$ in
Sect.~\ref{visibi} allows to compare $\ampav$ and $\amprad^{1/2}$
in more detail.

\begin{figure}
\includegraphics[width=8.88cm]{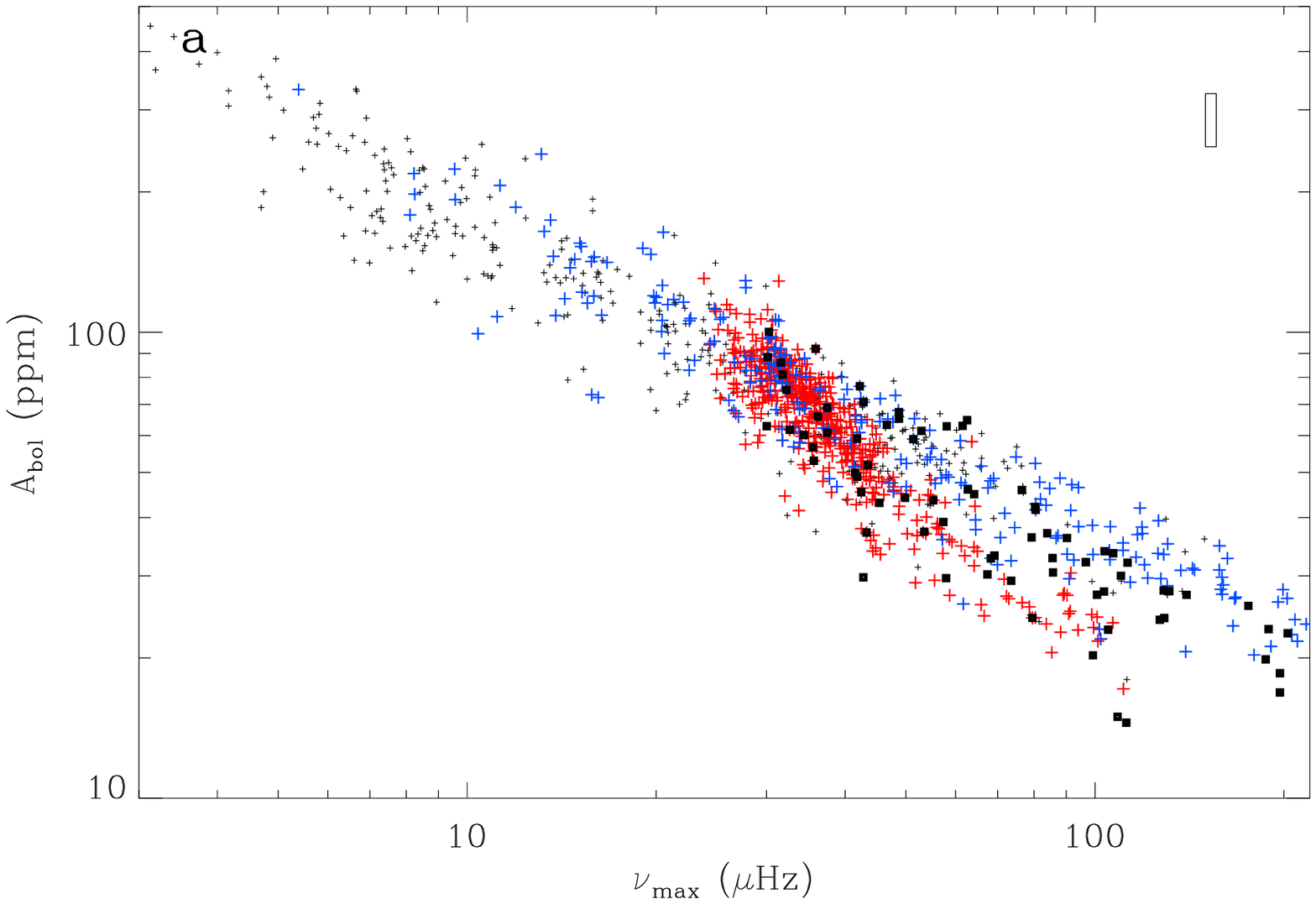}
\includegraphics[width=8.88cm]{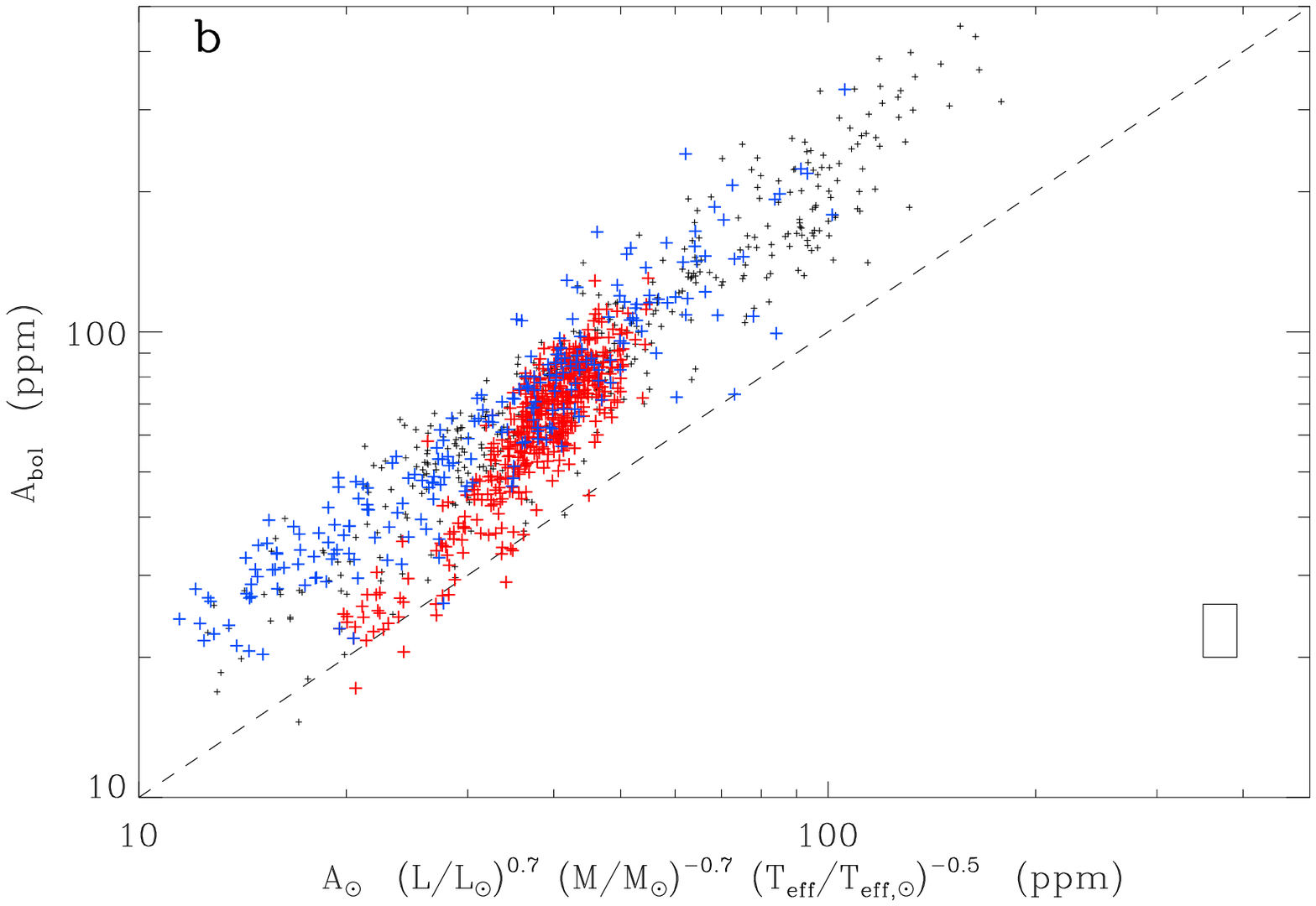}
\includegraphics[width=8.88cm]{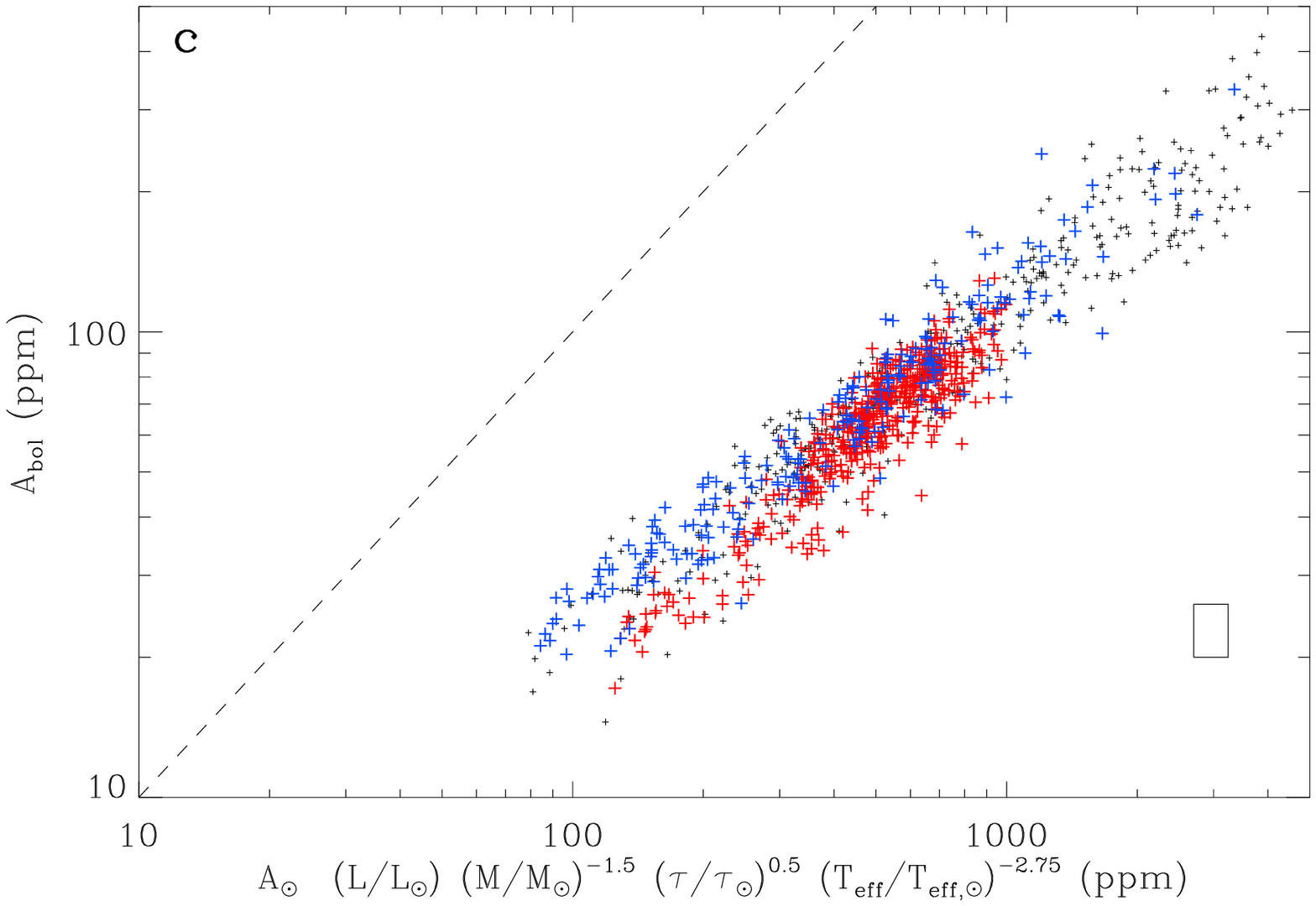}
\caption{Bolometric amplitude $\abol$, in ppm, as a function of
$\numax$ (panel {\bf a}), $A_\odot \, (LM_\odot/L_\odot M)^{0.7}
(\Teff / T_\odot)^{-0.5}$ (panel {\bf b}) and $A_\odot \,
(L/L_\odot) (M/M_\odot)^{-1.5} (\tauosc / \tau_\odot)^{0.5} (\Teff
/ T_\odot) ^{-2.75}$ (panel {\bf c}), with the solar value
$A_\odot \simeq 2.53$\,ppm. Evolutionary status and error bars are
indicated as in Fig.~\ref{correl-masse}. The dashed lines show the
1:1 correspondence.
\label{amplitude_LM}}
\end{figure}

\section{Bolometric amplitudes of radial modes\label{bolo}}

We now use the bolometric correction performed for \Kepler\
observations by \cite{2011A&A...531A.124B} to translate the
observed radial amplitudes to bolometric amplitudes. We then
examine how the bolometric radial amplitudes scale with various
parameters and test the available theoretical amplitude scaling
relations.

\subsection{Bolometric correction}

The bolometric correction provided by \cite{2011A&A...531A.124B}
allows us to derive bolometric amplitudes  from the radial
amplitudes:
\begin{equation}\label{correction_bolometrique}
   \abol = \amprad^{1/2} \ \corbol  \hbox{ \ with \ } \corbol = (\Teff / T_K)^{0.80}
\end{equation}
with $T_K = 5\,934$\,K. This correction accounts for the
wavelength dependence of the photometric variation integrated over
the \Kepler\ bandpass.  The variation of $\abol$ with $\numax$
shows clearly the differences between RGB and clump stars
(Fig.~\ref{amplitude_LM}a). We note that the difference is much
more pronounced for the secondary clump, which contains  objects
with higher mass than the primary clump.

\subsection{Comparison with $(L/M)^{0.7}$}

We have compared these amplitudes to different relations depending
on the stellar luminosity and mass. As in Sect.~\ref{fine}, both
$L$ and $M$ were derived from the usual asteroseismic scalings
relations. As discussed above, such scalings avoid the use of grid
modelling, which only works at fixed physics, but certainly lower
the quantitative validity of the outputs since the scaling
relations are not yet accurately calibrated.

We have tested the variation in $(L/M)^{0.7}$ proposed by
\cite{2007A&A...463..297S} for Doppler measurements, and verified
that the observed exponent is effectively near the value 0.7
derived from 3-D simulations. As already shown by
\cite{2010A&A...517A..22M} and \cite{2010ApJ...723.1607H}, we find
a slightly higher exponent, of about 0.8 (Fig.~\ref{amplitude_LM}b
and Table~\ref{table-amp}). We have introduced the correction from
Doppler to photometric measures, expressed by a factor
$\sqrt{\Teff}$, under the assumption that the oscillations
propagate adiabatically \citep{1995A&A...293...87K}. We also
adopted in the scaling relation the bolometric amplitude $A_\odot
\simeq 2.53$\,ppm of solar radial modes observed in broad-band
photometry \citep{2009A&A...495..979M}. We note that the relation
underestimates the amplitude by a factor of about 2, and that the
exponents are different for the different evolutionary states
(Fig.~\ref{amplitude_LM}b).

The theoretical relation proposed by \cite{2007A&A...463..297S}
was derived for main-sequence stars, and applying it to red giants
is a large extrapolation. Indeed, investigating mode driving in
red giants would require a non-adiabatic treatment.  From the
oscillation energy equation \citep[e.g.][Chap. IV, Eq.
21.14]{1989nos..book.....U}, dimensional arguments show that
non-adiabatic effects scale as the ratio $L/M$, which is roughly
50 times greater for red giants than for main-sequence stars.  We
might therefore expect the physics underlying mode driving to be
different for the two cases. In principle, non-adiabatic effects
must also be considered when converting between velocity
amplitudes (as provided by theoretical computations) and
photometric amplitudes (as measured with \Kepler).  However, in
the absence of a reliable non-adiabatic theory, we have to adopt
the adiabatic conversion
\citep[e.g.][]{1995A&A...293...87K,2010A&A...509A..15S}. These
points deserve thorough theoretical investigation, but this is
beyond the scope of this paper.

\subsection{Comparison with $L/M^{1.5}$}

We also tested the revised amplitude scaling relation suggested by
\cite{2011A&A...529L...8K}. It includes not only the mass and
luminosity, but also the effective temperature and the mode
lifetimes $\tauosc$:
\begin{equation}\label{amplKB}
    \abol \propto {\xAKB}
    \hbox{ \ with \ }
    {\xAKB} ={(L/L_\odot) \ (\tauosc / \tau_\odot)^{0.5} \over (M/M_\odot)^{1.5} \ (\Teff / T_\odot)^{2.75}}
    .
\end{equation}
\cite{2011A&A...529A..84B} explored the observed mode lifetime for
CoRoT data. They suggested that for red giants, $\tauosc$ is about
30 days and varies as $\Teff^{-0.3 \pm 0.8}$. We use these
estimates here. When expressed as a function of $\xAKB$, the two
branches corresponding to RGB and clump stars show a closer
correlation (Fig.~\ref{amplitude_LM}c), so that the distribution
of $\abol$ as a function of $\xAKB$ is less dispersed than the
distribution as a function of $(L/M)^{0.7}$. However, the best fit
gives a variation $\abol \propto \xAKB^{0.71\pm0.02}$ instead of
$\abol \propto \xAKB$ proposed by \cite{2011A&A...529L...8K}. As a
consequence, this theoretical relation largely overestimates the
observed bolometric radial amplitudes, as also shown by
\cite{2011arXiv1109.3460H} and \cite{2011ApJ...737L..10S}. This
means that the proposed relation does not provide a complete
physical explanation. We also note that the prediction of $\abol$
proportional to $L$ is observationally unlikely. For red giants,
$L$ is the most rapidly varying parameter in Eq.~\ref{amplKB}, so
that the observed slope in Fig.~\ref{amplitude_LM}b almost
directly translates into the luminosity exponent. The agreement
between observed and predicted values would require $\tauosc$ to
be approximately proportional to $\Dnu$, which is clearly not
observed \citep{2010ApJ...723.1607H,2011A&A...529A..84B}.

Despite the quantitative disagreement with theoretical
predictions, we see from Figs.~\ref{amplitude_LM}b and c that an
empirical scaling relation can describe the variation of $\abol$
as a function of the stellar luminosity, mass and effective
temperature. Such a fit based on red giants in open clusters,
hence with an accurate mass estimation, was proposed by
\cite{2011ApJ...737L..10S}, which agrees with a fit based on red
giants, subgiants and main-sequence stars performed by
\cite{2011arXiv1109.3460H}.

\begin{table}
\caption{Bolometric amplitudes}\label{table-amp}
\begin{tabular}{llcc}
\hline
scaling with&     & coefficient  & exponent \\
\hline
$\numax$ &all  & $ 860\pm30 $ & $-0.71\pm0.01$ \\
         &clump& $4700\pm500$ & $-1.18\pm0.03$ \\
         &RGB  & $ 650\pm60 $ & $-0.63\pm0.02$ \\
$L/M$    &all  & $ 3.5\pm0.2$ & $ 0.79\pm0.01$ \\
         &clump& $0.92\pm0.11$& $ 1.12\pm0.04$ \\
         &RGB  & $ 4.8\pm0.4$ & $ 0.71\pm0.02$ \\
$ \xAKB$ &all  & $ 8.2\pm0.2$ & $ 0.70\pm0.01$ \\
         &clump& $ 5.3\pm0.3$ & $ 0.83\pm0.02$ \\
         &RGB  & $ 9.8\pm0.4$ & $ 0.65\pm0.02$ \\
\hline
\end{tabular}
\end{table}

\begin{figure}
\includegraphics[width=8.8cm]{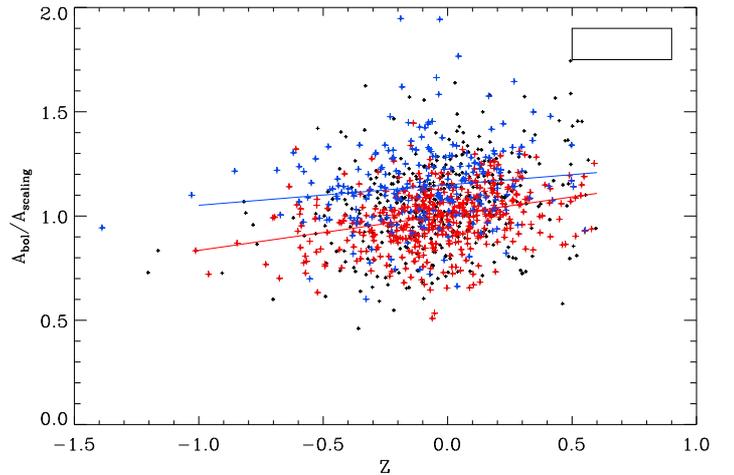}
\caption{Variation in metallicity of the bolometric amplitude
scaled to a best fit in $L/M^{1.5}$. Same color code as in
Fig.~\ref{correl-masse}. The red and blue lines corresponds to the
linear fits in $Z$. The uncertainties on $Z$ are obtained from
\cite{2011AJ....142..112B}.\label{metal}}
\end{figure}

\begin{figure}
\includegraphics[width=8.8cm]{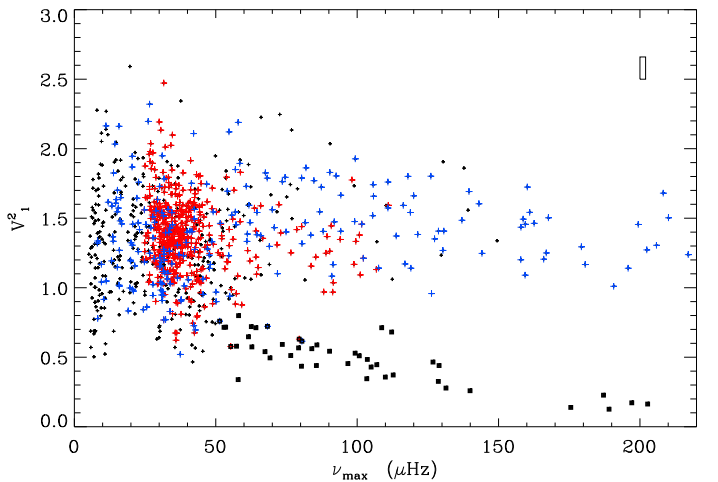}
\caption{Visibility $\visiun$ as a function of $\numax$, with the
same color code as in Fig.~\ref{correl-masse}. Large black symbols
indicate the population of stars with very low $\visiun$ values.
\label{V1_numax}}
\end{figure}

\begin{figure}
\includegraphics[width=8.8cm]{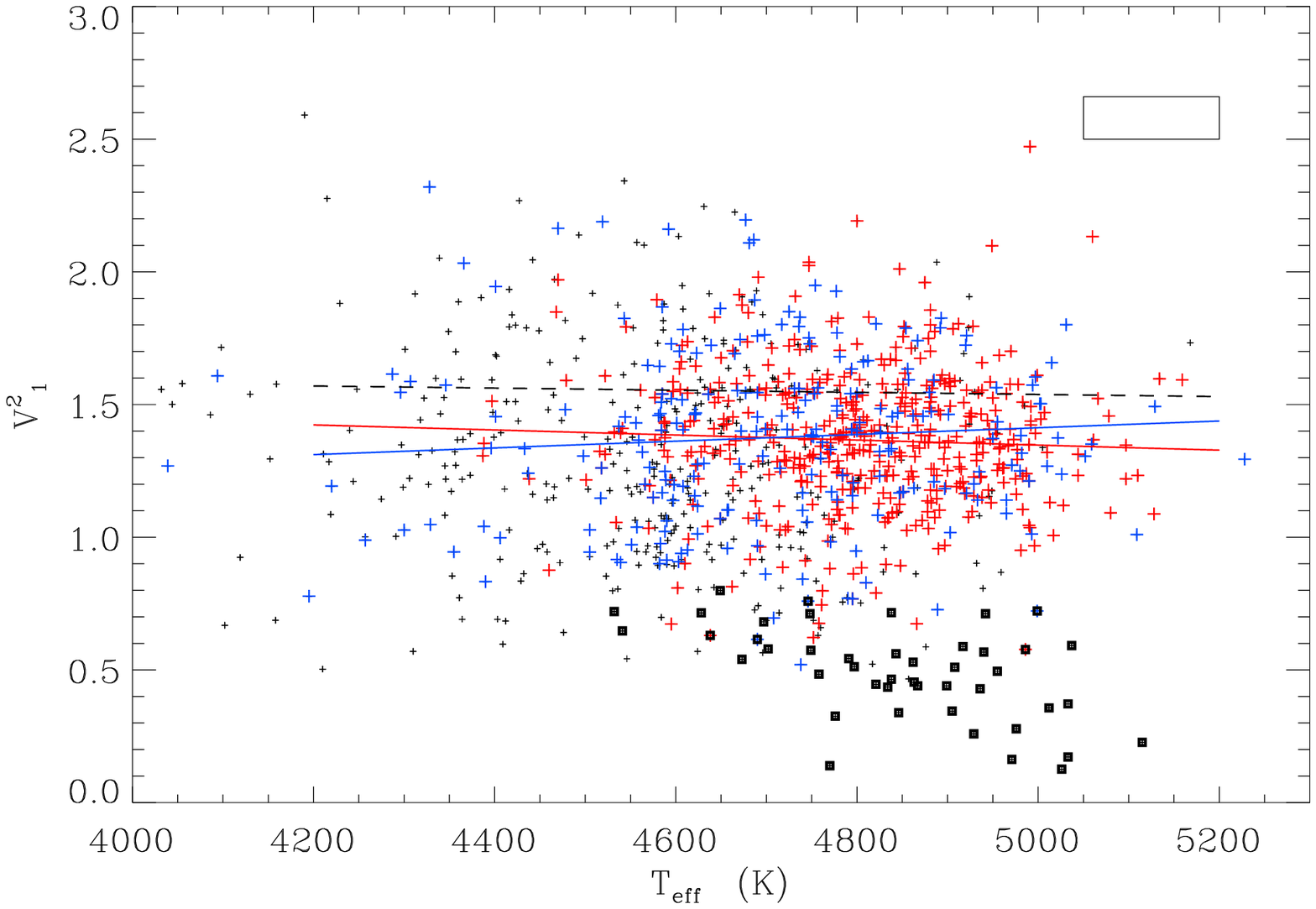}
\includegraphics[width=8.8cm]{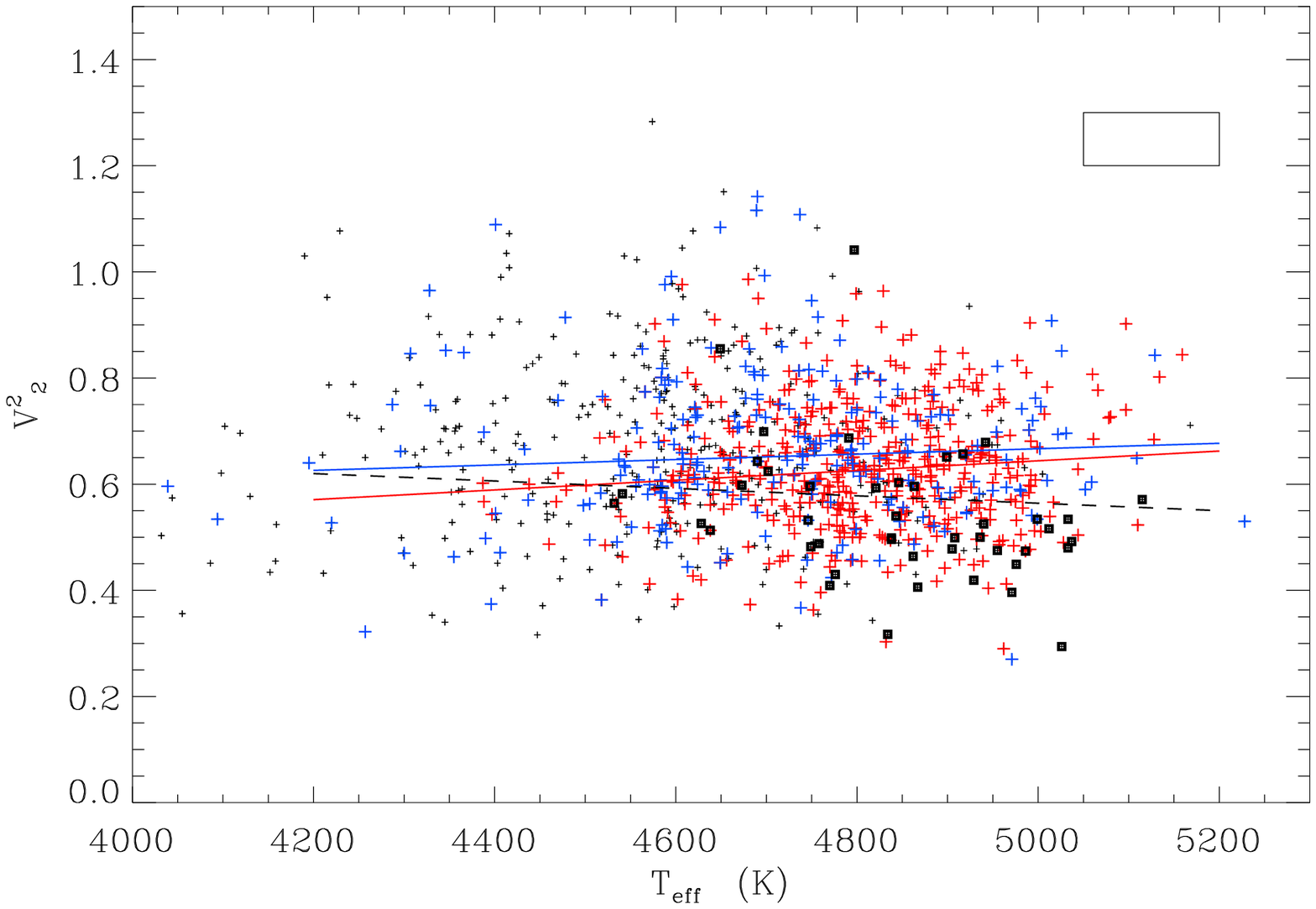}
\includegraphics[width=8.8cm]{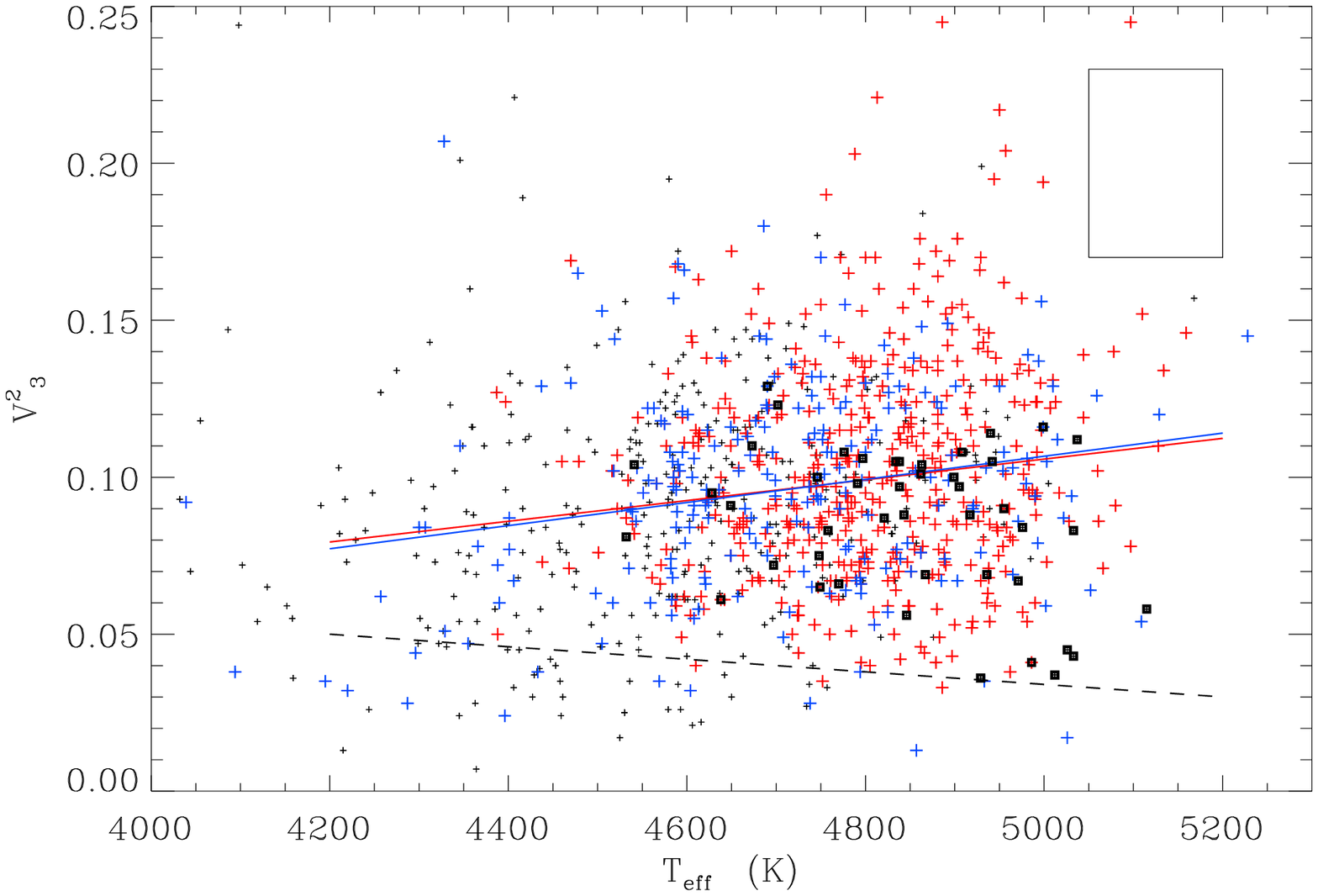}
\caption{Visibility of the $\ell=1$, 2 and 3 modes (from top to
bottom), as a function of the effective temperature. Same color
code as in Fig.~\ref{V1_numax}. Dashed lines are the fits derived
from \cite{2011A&A...531A.124B}. Blue and red solid lines are the
fits for RGB and clump stars, respectively. \label{visibT}}
\end{figure}

\begin{figure}
\includegraphics[width=8.8cm]{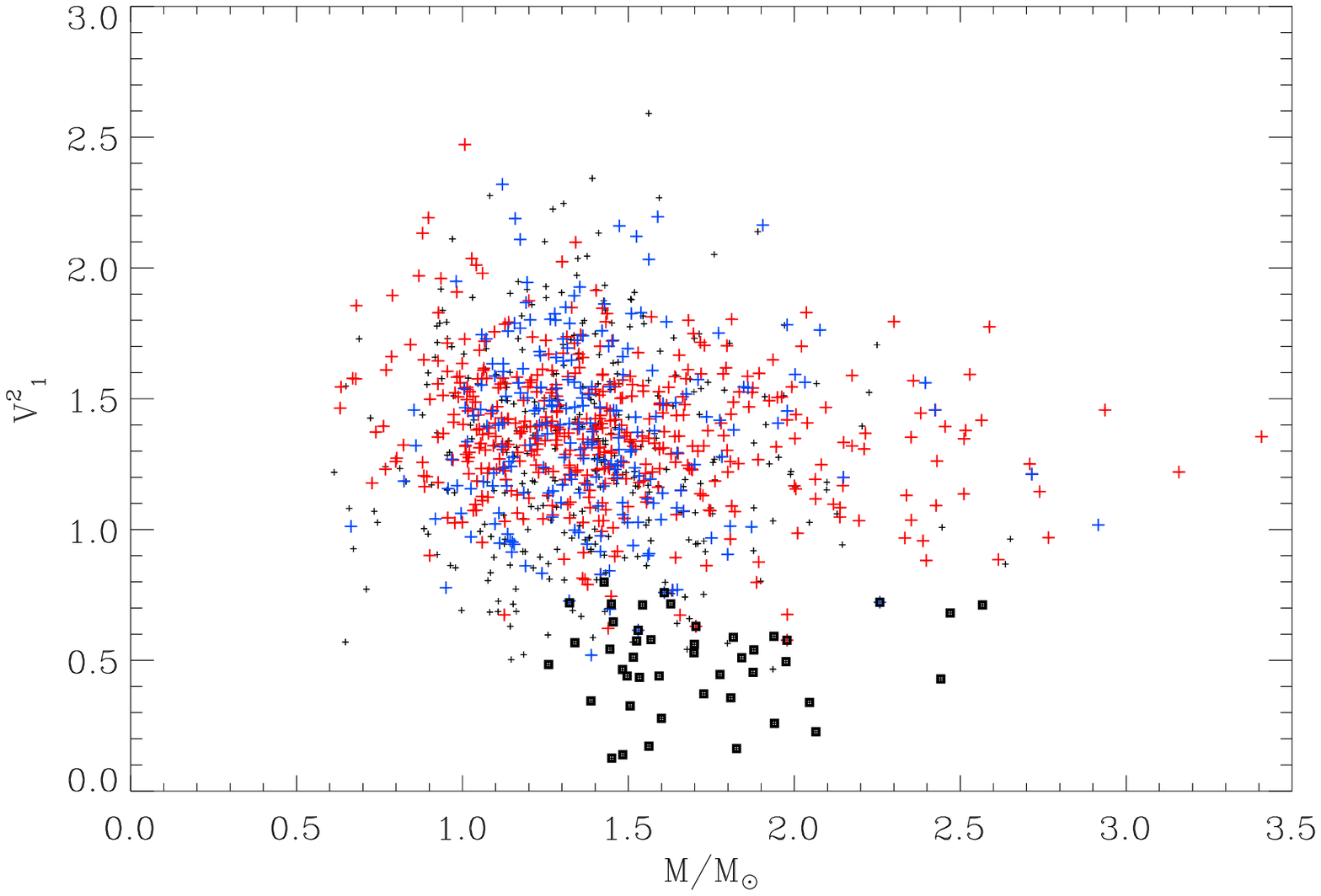}
\includegraphics[width=8.8cm]{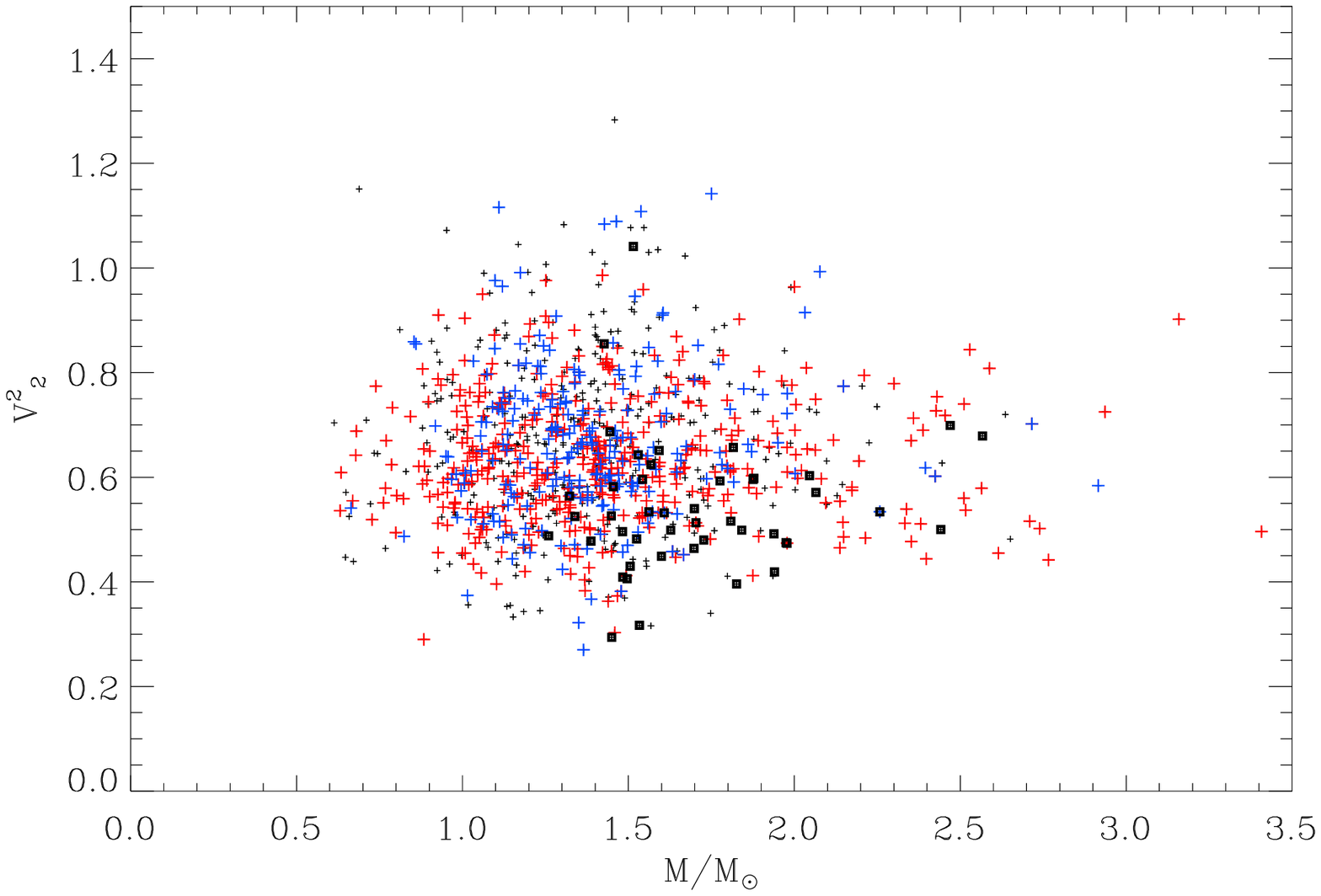}
\caption{Visibility of the $\ell=1$ and 2 modes, as a function of
the seismic mass. Same color code as in
Fig.~\ref{V1_numax}.\label{visibM}}
\end{figure}

\subsection{Influence of metallicity}

It has been suggested that oscillation amplitudes depend on the
stellar metallicity
\citep{1999A&A...351..582H,2010A&A...509A..15S}. In order to test
this dependence, we have plotted the bolometric amplitudes as a
function of the metallicity (Fig.~\ref{metal}). This comparison
requires normalized amplitudes, which were obtained using a fit in
$L/M^{1.5}$ \citep{2011arXiv1109.3460H}. The metallicity values
were taken from the \Kepler\ Input Catalog
\citep{2011AJ....142..112B} and have uncertainties of about
0.4\,dex. We found that an increase of 1 dex in $Z$ gives a
moderate amplitude increase of about 10$\pm$5\,\%. Further work,
based on improved values of the metallicity, will be necessary to
make the link more precise.

\section{Spatial response functions\label{visibi}}

We have examined the visibility of each degree, obtained from
\begin{equation}\label{calculV}
    \visib_\ell = \langle A^2_\ell \rangle\ / \ \langle A^2_0 \rangle
    ,
\end{equation}
with $\langle A^2_0\rangle$ given by Eqs.~\ref{calculA} and
\ref{calculAmoy}, and $\langle A^2_\ell\rangle$ obtained in a
similar way.
For pure p modes in a 4800 K red giant, one expects spatial
response functions in power to vary as $\visiun \simeq 1.54$,
$\viside \simeq 0.58$ and  $\visitr \simeq 0.043$. These values,
derived from \cite{2011A&A...531A.124B}, assume energy
equipartition between the different degrees and take into account
the influence of the limb darkening. They are expected to decrease
when the effective temperature increases (Table \ref{tab_visi}).
Fig.~\ref{difference} shows the background-corrected spectra of
different red giants exhibiting different mode visibilities.

\subsection{Dipole modes}

The visibilities $\visiun$ of the dipole modes are plotted in
Fig.~\ref{V1_numax} as a function of $\numax$ and in
Fig.~\ref{visibT}a as a function of the effective temperature. We
first note a large dispersion of the values and a discrepancy
between observed and predicted values. In general, the observed
$\visiun$ appear to be smaller than expected. Despite including
the mixed modes in a very broad frequency range, in fact larger
than $\Dnu/2$, we observed $\langle\visiun\rangle = 1.34\pm0.02$
instead of the expected value of 1.54 (Table \ref{tab_visi}).
Individual uncertainties on $\visiun$ are about $\pm 0.08$. They
are principally due to the background modelling and to the
difficulty of automatically identifying all mixed modes.


\subsubsection{Very weak $\visiun$ values\label{veryweak}}

We note the presence of a family of stars with a very weak, almost
absent $\ell=1$ oscillation pattern. It mostly comprises less
evolved stars, with $\numax \ge 50\,\mu$Hz and $\visiun\le 0.8$,
which appear clearly as outliers. As a consequence of the low
$\ell=1$ amplitudes, their evolution status is not determined,
except for four stars, two in the RGB and two in the clump. Very
low $\visiun$ values are also measured at the clump, but without
the clear difference compared to other stars. Therefore, we do not
include them among the set of stars with very weak $\visiun$
values. We limit it to $\numax \ge 50\,\mu$Hz and a low $\visiun$
(about $12.5\; \numax^{-0.72}$, with $\numax$ in $\mu$Hz).

Different tests have shown that, apart from a low $\ell=1$
visibility, these stars mostly share the common characteristics of
other stars. Unsurprisingly, they have lower heights than expected
from the empirical scaling relation since $\ell=1$ modes are
almost absent, but their bolometric amplitudes are close to the
mean values. More precisely, if one believes that the location in
the $\abol(\numax)$ relation (Fig.~\ref{amplitude_LM}) makes it
possible to identify the evolutionary status, clump stars with a
low $\visiun$ show normal bolometric amplitudes, but RGB stars
with a low $\visiun$ have lower amplitudes than expected. This
could be related to the fact that these stars have slightly higher
masses compared to the mean values (Fig.~\ref{visibM}a). However,
some stars do present simultaneously a high asteroseismic mass and
a normal $\visiun$. Their $\ell=2$ visibilities are slightly lower
than average, but not significantly so. Furthermore, these stars
are uniformly distributed in temperature and metallicity.
Therefore, we have to conclude that only $\ell=1$ modes are
affected. One may imagine that their suppression results from a
very efficient coupling between pressure and gravity waves. This
efficient coupling yields a very high mode mass, hence a very low
observed amplitude at the surface. Examining the causes of very
low $\visiun$ values will need further work. It may also help to
understand the non-identification of $\ell=1$ modes in previous
observations of high-mass giants
\citep{2002A&A...394L...5F,baudin}.

\subsubsection{A surprising energy equipartition}

Having identified the population with abnormally low $\visiun$
values allows us to calculate the mean $\ell=1$ visibility of
`normal' red giants: it remains lower than the expected value
(1.38, versus 1.54), and uncertainties are unable to explain the
discrepancy. We note that the lower-than-expected $\visiun$ values
are independent of $\numax$, hence independent of the evolution.
Coming back to the way the amplitudes were determined, we see that
the sum of the squared amplitudes of all $\ell=1$ mixed modes
corresponding to a fixed radial order is only slightly less than
the total expected squared amplitude. We show later that the
deficit can be explained by the energy being spread far away from
the expected location of the $\ell=1$ p modes, which are not being
detected by the automated method presented in Sect.~\ref{ampli}.

In terms of coupling, the energy injected in the pressure waves
near the stellar surface seems to be shared among all mixed modes
associated to a given radial order. The energy equipartition seems
valid when all squared amplitudes of the mixed modes associated to
a given radial order are summed. Therefore, in terms of mode mass,
one derives the equivalence between the radial and non-radial
modes:
\begin{equation}\label{modemass}
    {1\over \modemass (n,0)} \simeq \sum_{\ng}  {1\over \modemass_{\ng}
    (n,1)}
\end{equation}
with $\ng$ being the gravity order of the g modes mixed with the p
mode of radial order $n$. Each mixed-mode mass $\modemass_{\ng}$
is much higher than the radial mode mass, as discussed by
\cite{2009A&A...506...57D}, but the total as defined by
Eq.~\ref{modemass} is close to the radial mode mass. Future work
will determine whether this is a coincidence or based on physical
principles.

\subsubsection{Variation of $\visiun$ with $\Teff$, $M$ and metallicity}

The variation of $\visiun$ with temperature matches the
expectation, although with large uncertainties
(Fig.~\ref{visibT}a, Table \ref{tab_visi}). The visibility
$\visiun$ is supposed to increase towards low temperatures. If the
extrapolation is valid, one may expect the observed  $\ell=1$
modes to be more dominant for red giants towards the tip of the
red giant branch. For these stars, the oscillation spectra cannot
yet be analyzed since the frequency resolution is not high enough.

We have also examined how the spatial responses $V_\ell^2$ vary
with the stellar mass (Fig.~\ref{visibM}a). We remark that the
high values of $\visiun$ are concentrated at intermediate mass,
between 1 and 1.7 times the solar mass. The visibility of $\ell=1$
modes is, on average, smaller for higher mass. The variation of
$\visiun$ with metallicity shows no significant correlation, but
this result may be due to the very uncertain metallicity
determinations in the \Kepler\ Input Catalog.

Since correlations between $\visiun$ and any fundamental stellar
parameters or seismic global parameters just discussed are not
very strong, we suspect that the scatter in $\visiun$ is related
to the conditions that govern the coupling of pressure and gravity
waves that contribute to the mixed modes. These conditions appear
to be highly sensitive to the locations of the inner and outer
cavities where g and p waves propagate, respectively. Future
analysis will have to show if one can use the differences in
visibility by correlating them with properties of the
eigenfrequency spectrum and drawing conclusions about the interior
structure.

\begin{table}
\caption{Mode visibility}\label{tab_visi}
\begin{tabular}{lccc}
\hline visibility & population & value at 4\,800\,K& $10^3 \ \diff\visib_\ell / \diff \Teff$\\
 \hline
$\visiun$ &expected&  1.54          & $-0.06$\\
          &clump& 1.34$\pm$0.02 & $-0.07\pm0.09$ \\
          &RGB  & 1.35$\pm$0.04 & $-0.10\pm0.12$\\
\hline
$\viside$ &expected& 0.58 & $-0.07$\\
          &clump& 0.59$\pm$0.01 & $-0.02\pm$0.06\\
          &RGB  & 0.64$\pm$0.02 & $-0.03\pm$0.06\\
\hline
$\visitr$ & expected & 0.036 & $-0.02$\\
          &clump& 0.056$\pm$0.01 &$+0.03\pm$0.01\\
          &RGB  & 0.071$\pm$0.01 &$+0.04\pm$0.01\\
\hline
$\sum_{\ell=0}^3 \visib_\ell$ & expected & 3.16 & $-0.15$\\
          &clump& 2.98$\pm$0.02& $-0.07\pm$0.10\\
          &RGB  & 3.08$\pm$0.04& $-0.11\pm$0.15\\
\hline
\end{tabular}
\end{table}

\subsection{Quadrupole modes}

The visibilities $\viside$ of the quadrupole modes are plotted in
Fig.~\ref{visibT}b. Compared to dipole modes, the dispersion is
much lower. Uncertainties on $\viside$ are smaller, about $\pm
0.05$, since the integration interval is reduced, and since
uncertainties in the background modelling have less influence.

The mean value is very close to the expected value. This is
surprising, since $\ell=2$  modes are also mixed and so, as for
dipole modes, one would expect lower $\viside$ values. However,
with a narrower g-mode spacing than for $\ell=1$ modes, in
agreement with the asymptotic expectations, the density of
$\ell=2$ mixed modes is much larger than for $\ell=1$. Perhaps
this larger density, or a better coupling, compensates for the
larger mode mass. This agrees with the better trapping of
quadrupole modes, compared to dipole modes, expected by
\cite{2009A&A...506...57D}.

We have checked that the variation of $\viside$ with the effective
temperature also agrees with the theoretical expectation. Contrary
to $\visiun$, the mass dependence of $\viside$ seems to be nearly
flat (Fig.~\ref{visibM}b).

\subsection{$\ell = 3$ modes}

The visibilities of $\ell=3$ modes have been calculated in a
similar way (Fig.~\ref{visibT}c). Absolute uncertainties on
$\visitr$ are smaller than for $\visiun$ and $\viside$, about $\pm
0.03$, but relative uncertainties are much higher due to their low
intrinsic amplitudes. Individual checks confirm the reality of
$\ell=3$ peaks with large amplitudes. However, we also identify
the spurious contribution of g-dominated $\ell=1$ mixed modes,
which are located far away from the expected position of pure
$\ell=1$ pressure modes. They can contribute a significant part of
the energy automatically assigned to the $\ell=3$ modes. This may
explain part of the deficit of the $\ell=1$ energy reported above.

We note that the vast majority of $\ell=3$ modes are observed as
single narrow peaks. This should indicate that, most of the time,
only one $\ell=3$ mixed mode per $\Dnu$-wide frequency interval is
efficiently excited. Further observations with a higher frequency
resolution will help to make this point more clear.

\begin{figure}
\includegraphics[width=8.8cm]{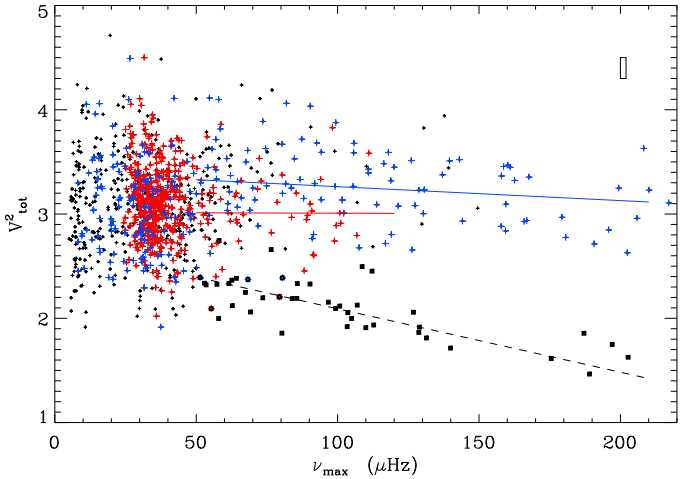}
\caption{Visibility $\visitot$ as a function of $\numax$. Same
color code as in Fig.~\ref{V1_numax}. The fits of RGB, clump and
low $\visiun$ stars, in blue and red solid lines or black dashed
line, respectively, are computed for $\numax\ge 50\,\mu$Hz.
\label{Vtot_numax}}
\end{figure}

\begin{figure}
\includegraphics[width=8.8cm]{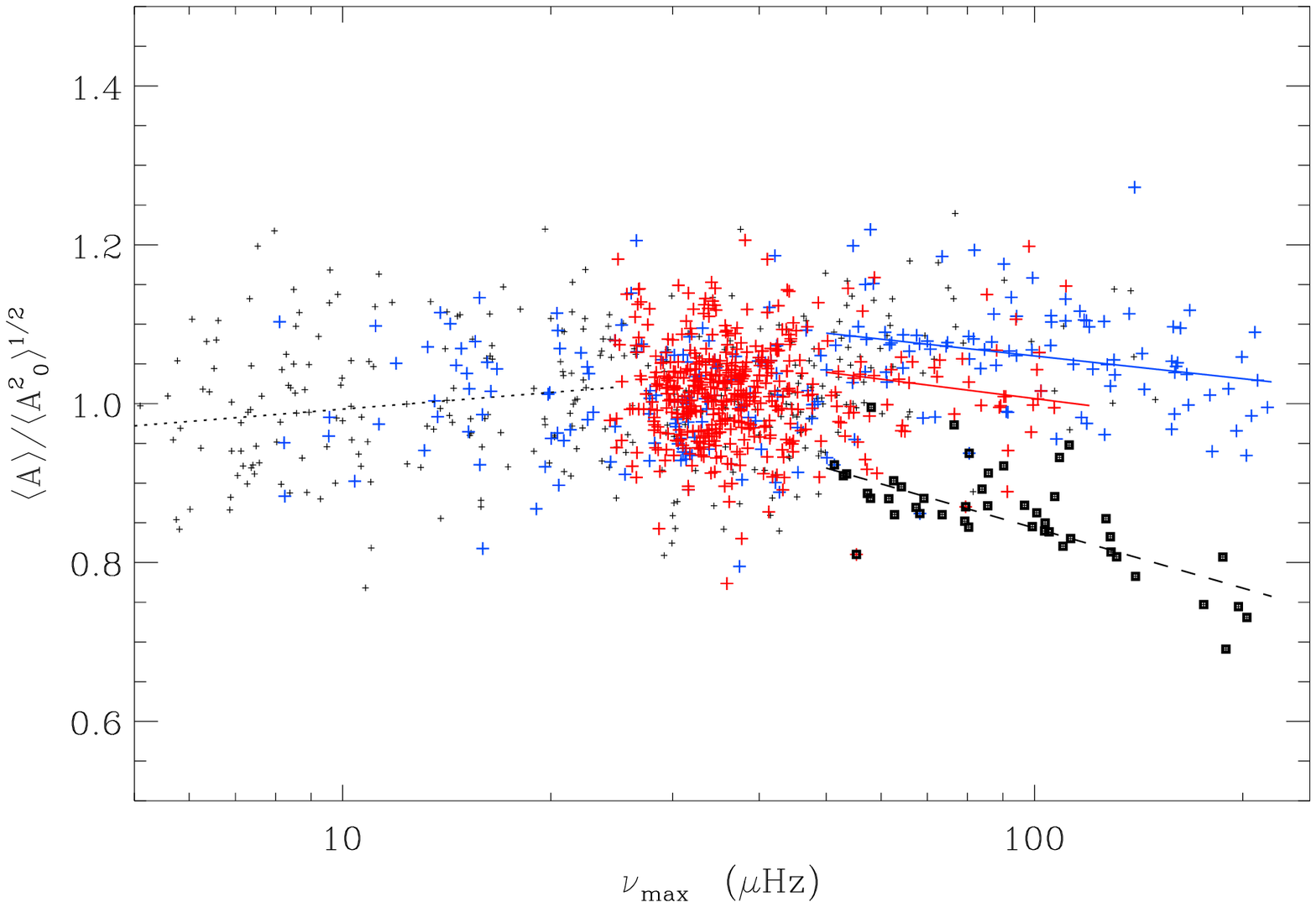}
\caption{Ratio $\ampav / \amprad^{1/2}$ as a function of $\numax$.
Same color code as in Fig.~\ref{V1_numax}. The fits of RGB, clump
and low $\visiun$ stars, in blue, red, and black, respectively,
are valid at $\numax\ge 50\,\mu$Hz. The dotted line indicates the
fit at frequency lower than 30\,$\mu$Hz. \label{compar_amprad}}
\end{figure}

\subsection{Total visibility and radial amplitudes}

We have estimated the total visibility $\visitot = \sum_{\ell=0}^3
\visib_\ell$ and note a large spread of the values, with RGB and
clump stars having different mean values (Fig.~\ref{Vtot_numax}).
We also note that the stars with a low $\visiun$ also show a low
total visibility. \cite{2011A&A...531A.124B} have calculated red
giant limb-darkening functions, and derived a mean value $\visitot
\simeq 3.18$ for $\Teff = 4\,800$\,K and solar metallicity. We
find a mean value of about 3.06, in better agreement with the
value 3.04 calculated by \cite{2010ApJ...713L.176B}.

The mean value of $\visitot$ can be used to improve the
coefficient defining the globally averaged amplitude $\ampav$
given by Eq.~\ref{average}. However, this globally averaged value
is only indicative, due to the dispersion in $\visitot$
(Fig.~\ref{Vtot_numax}). This limits the use of $\ampav$ for
deriving precise red giant radial amplitudes. Therefore, we
discourage using the globally averaged value $\ampav$ for
estimating the bolometric amplitude of a given red giant. Using
$\ampav$ instead of $\amprad^{1/2}$ for establishing scaling
relations is possible, but with an uncertainty on the exponent in
$L$ of about 0.04. This value results from the differences in the
slopes observed in the scaling relation of $\ampav /
\amprad^{1/2}$ to $\numax$ (Fig.~\ref{compar_amprad}), taking into
account the fact that $L \propto \numax$.

\section{Conclusion\label{conclusion}}

With \Kepler\ photometric data recorded up to Q7, providing
590-day long time series, we have determined scaling relations of
the global parameters that describe the oscillation power excess
of red giants. About 1200 red giants were analyzed.

\subsection{Global parameters and scaling relations}

We have compared different methods, and shown that the background
modelling influences the results. This influence does not prevent
firm conclusions, but certainly needs further work to fully
disentangle the fit of the granulation background from the
oscillation power excess. There are indications that Harvey
components with an exponent of 2 are sufficient for modelling the
background of red giant spectra in the frequency range around
$\numax$.

We have found a significant influence of the red giant
evolutionary status on the scaling relations describing the
asteroseismic and fundamental stellar parameters. Following
\cite{2011A&A...532A..86M}, we show that red-clump stars have a
mass distribution clearly linked to the seismic parameter
$\numax$, hence to the stellar radius. Therefore, the scaling
relations expressed as a single power law of $\numax$ present a
large dispersion around the clump.

As a consequence of the different mass distributions, red giants
on the RGB have larger oscillation amplitudes than the clump
stars. Members of the secondary clump have different properties:
compared to RGB stars, they have much higher masses, lower
oscillation amplitudes, but with more modes excited.

We have compared the energy in the oscillations to its counterpart
in the stellar background. The mean height in the oscillations is
a fixed fraction of the energy density in the background at
$\numax$. This fraction depends only slightly on the evolutionary
status. Since both granulation and oscillation are due to
convection, such a result clearly indicates that the excitation
mechanism of the oscillation has the same efficiency for a very
large variety of red giants.

For both the background height and the bolometric amplitude, the
agreement between the RGB and clump stars is better when
considering in the scaling relations a mass dependence not exactly
opposite to the luminosity dependence. The background height seems
to scale as $L^2/M^3 \Teff^{5.5}$, as predicted by
\cite{2011A&A...529L...8K}.

\subsection{Bolometric amplitudes and mode visibility}

We have proposed a new method for deriving radial and non-radial
amplitudes, and we have shown that the amplitudes obtained with a
global integration are not appropriate for red giants, due to the
presence of mixed modes that can significantly modify the total
mode visibility.

Radial amplitudes have been translated into bolometric amplitudes
for comparison with theoretical expectations. We have shown that
current theoretical scaling relations are unable to provide an
acceptable fit to the observed photometric amplitudes. There are
strong observational indications in favor of an exponent in
luminosity in the range [0.7, 0.8]. A negative exponent of the
stellar mass larger, in absolute value, than the luminosity
exponent helps to reduce the discrepancy between RGB and clump
stars \citep{2011arXiv1109.3460H,2011ApJ...737L..10S}, but it
seems that ingredients other than $L$ and $M$ are necessary to
properly account for the observations.

We have also derived the mode visibility of $\ell$=1, 2 and 3
modes. The presence of mixed-modes reduces the observed values
compared to expectations. As a result, we observe a new type of
energy equipartition, where the squared amplitude of the pure
pressure modes (as radial modes) seems to be shared by all mixed
modes. We note a high dispersion that cannot be related to global
parameters, and have identified a class of objects with very low
$\ell=1$ amplitudes. It probably comprises red giants more massive
than $1.3\,M_\odot$ where the coupling between pressure and
gravity waves is so efficient that all $\ell=1$ mixed modes have a
very high mode mass.

\begin{appendix}
\section{Normalization of the power density spectra\label{appendix-normalisation}}

\begin{table*}
\caption{Characteristics of the different methods
\label{pipelines}}
\begin{tabular}{lp{2.3cm}p{2.3cm}p{2.3cm}p{2.3cm}p{2.9cm}}
\hline
pipeline  & A2Z$^a$& CAN$^b$&  COR$^c$& OCT$^d$& SYD$^e$\\
\hline
spectrum    &\multicolumn{5}{c}{\dotfill\ Lomb-Scargle periodogram $\to$ power density spectrum (PDS)\dotfill}\\
frequency axis&\multicolumn{5}{c}{\dotfill\ corrected for the duty cycle \dotfill}\\

frequency sampling&\multicolumn{5}{c}{\dotfill\ $\delta \nu = 1/T$, with $T=N\;\LC$ the total observation duration  \dotfill}\\
          &\multicolumn{5}{c}{\dotfill\ $\fnyquist = 0.5 / \LC$, with
$\LC$ the \Kepler\ long-cadence sampling
            \dotfill}\\
PDS normalization&\multicolumn{5}{c}{\dotfill\ $\sigma_\nu^2 = 2\, \sigma_t^2 \ \LC/N$\dotfill}\\
\hline
$\numax$  &\multicolumn{5}{c}{\dotfill\ centroid Gaussian \dotfill}\\
          &        &         &         &       & or peak of the smoothed
          spectrum\\
smoothing &1$-\Dnu$ Gaussian filter & none & \multicolumn{3}{c}{\dotfill\ 1$-\Dnu$ Gaussian filter \dotfill}\\
\hline
background&global & global   &  local  & local & global  \\
          &Harvey & Harvey   & $\propto \nu^{-\beta}$ &second order polynomial& Harvey modified \\
value of exponent & free        & 4       &  free        & 2 & 2
and 4\\
number of components&2      &  3       &       1 & 1       & 2  \\
 \hline
\end{tabular}

References: $a)$ \cite{2010A&A...511A..46M}; $b)$
\cite{2010A&A...522A...1K}; $c)$ \cite{2009A&A...508..877M}; $d)$
\cite{2010MNRAS.402.2049H}; $e)$ \cite{2009CoAst.160...74H}. For
more information on the modelling of the background, see
\cite{2011arXiv1109.1194M}.

\end{table*}

Different methods have been used and compared for the analysis of
the power excess
\citep{2009CoAst.160...74H,2009A&A...508..877M,2010MNRAS.402.2049H,2010A&A...522A...1K,2010A&A...511A..46M}.
A comprehensive comparison of the properties and characteristics
of the methods was given by \cite{2011A&A...525A.131H}, but only
dealt with the large separation $\Dnu$ and the frequency $\numax$
of maximum oscillation signal. A comparison of complementary
analysis methods applied to the \Kepler\ short-cadence data by
\cite{2011MNRAS.tmp..892V} presented results on the maximum mode
amplitude, but not for red giants. For the current work on red
giants, parameters characterizing the oscillation power excess
have been compared.

To ensure a correct comparison of results obtained with different
methods, it was first necessary to normalize the outputs. The
computation of power density spectra performed with Lomb-Scargle
periodograms has taken into account the correction for the duty
cycle of each target, in order to obtain spectra with the
frequency resolution corresponding to the total observation time
and frequency up to the Nyquist frequency ($\fnyquist = 283.2\
\mu$Hz) related to the mean time sampling $\LC$ of the \Kepler\
long-cadence data. We chose to compute power density spectra (PDS)
with the following normalization: a white noise signal recorded at
the \Kepler\ long-cadence sampling $\LC$ with a noise level
$\sigma\ind{t}$ (in ppm) gives a PDS with a spectral density
$\sigma_\nu$, such that $\sigma^2_\nu = 2 \sigma^2\ind{t} \LC / N$
where $N$ is the number of points in the time series. The
characteristics of the methods used for this work are briefly
described in Table~\ref{pipelines}.

\begin{figure}
\includegraphics[width=8.88cm]{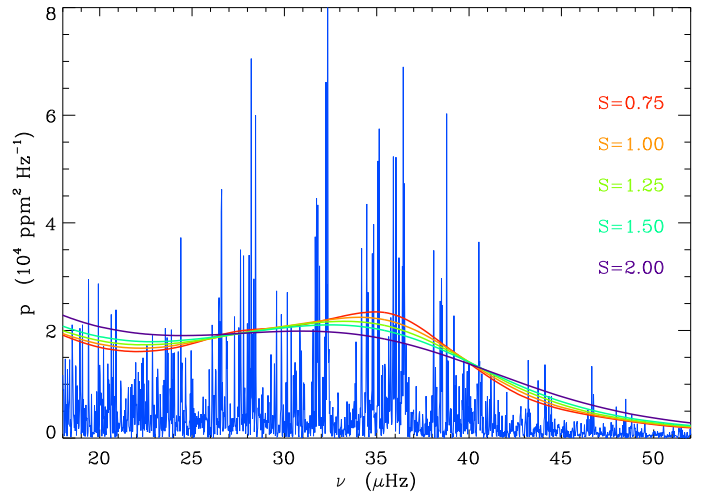}
\caption{Influence of the width of the smoothing parameter for a
typical red-clump giant (KIC 1161618). The different colors
indicate the different values of the smoothing (in units of
$\Dnu$).\label{fig-smoothing}}
\end{figure}

\begin{table}
\caption{Influence of the smoothing on the seismic global
parameters of KIC 1161618.}\label{table-smoothing}
\begin{tabular}{cccccc}
\hline
$\smoo$                   & 0.75 &    1        & 1.25 & 1.5  & 2 \\
\hline
$\numax(\smoo)\,/\,\numax(1)$ & 1.00 & \emph{1.00} & 0.98 & 0.97 & 0.96 \\
$\dnuenv(\smoo)\,/\,\dnuenv(1)$&1.00 & \emph{1.00} & 1.05 & 1.11 & 1.17 \\
$\Hmax(\smoo)\,/\,\Hmax(1)$   & 1.00 & \emph{1.00} & 0.89 & 0.84 & 0.74 \\
$\Bmax(\smoo)\,/\,\Bmax(1)$   & 0.99 & \emph{1.00} & 1.10 & 1.15 & 1.28 \\
HBR$(\smoo)$\,/\,HBR$(1)$     & 1.01 & \emph{1.00} & 0.81 & 0.73 & 0.58 \\
$\expB (\smoo)\,/\,\expB (1)$ & 0.99 & \emph{1.00} & 1.10 & 1.13 & 1.16 \\
\hline
\end{tabular}
\end{table}

\section{Smoothing\label{appendix-smoothing}}

Most of the methods use a smoothed spectrum to obtain the global
parameters of the Gaussian envelope (Sect.~\ref{global}). Since
the width of the Gaussian envelope of red giant oscillation
(Eq.~\ref{bosse}) is narrow \citep{2010A&A...517A..22M}, this step
must be performed carefully. An example for a typical red-clump
star is given in Fig.~\ref{fig-smoothing}. The parameters $\Hmax$
and $\Bmax$ have been calculated for the local description of the
background smoothed with different filter sizes and are summarized
in Table~\ref{table-smoothing}. We have used a Gaussian window for
performing the smoothing and have tested different values $\smoo$
of the FWHM. The mean value of the large separation acts as a
natural characteristic frequency, so we expressed the FWHM in
units of $\dnumoy$ and have tested values varying from
$0.75\,\dnumoy$ to $2.0\,\dnumoy$. When this value increases, the
measured value of $\numax$ decreases, with variations as large as
4\,\%, depending on the method. In parallel, $\Hmax$ decreases and
$\Bmax$ increases when the filter width increases, which results
in variations of the height-to-background ratio (HBR) larger than
a factor of two. As expected, $\dnuenv$ also increases with
$\smoo$. This variation can be easily understood by recognizing
that increasing the smoothing will spread the power of the
oscillations relative to the background. We consider an acceptable
compromise to be a  Gaussian filter with a FWHM equal to the mean
large separation $\dnumoy$. Such a width is large enough for
smoothing the influence of individual contributions of different
degrees, and narrow enough to avoid the dilution of the global
parameters, as shown by the example given in Table
\ref{table-smoothing}: when the parameter $\smoo$ gets larger than
1, all terms show significant variations.

This study shows the importance of the smoothing parameter. A
systematic variation induced by the smoothing as large as 4\,\% in
$\numax$ is important because this parameter plays a key role in
scaling relations for seismic parameters
\citep{2010A&A...517A..22M, 2010ApJ...723.1607H}. Since $\numax$
is used for deriving the seismic mass and radius
\citep{2010A&A...517A..22M,2010A&A...522A...1K,2011MNRAS.414.2594H},
a bias should be avoided in order to avoid subsequent biases in
the stellar parameters. A systematic relative error
${\sigma}_{\numax}$ on $\numax$ translates into a bias in the
determination of the derived stellar parameters, of the order of
${\sigma}_{\numax}$ and $3\,{\sigma}_{\numax}$ for, respectively,
the relative precision of the radius and mass.

A variation of a factor of two of the HBR depending on the
smoothing means that studying the partition of energy between the
oscillation and the background requires a careful description.
According to the solar example, longer time-series are certainly
required to reach the necessary precision to draw firm
conclusions.

\section{Mode identification\label{appendix-identification}}

The determination of the mode visibility (Sect.~\ref{fine} and
\ref{visibi}) requires the complete identification of the p-mode
spectrum. This is automatically given by the red giant universal
oscillation pattern \citep{2011A&A...525L...9M}, with the
parametrization of the dimensionless factor $\varepsilon$ of the
asymptotic development \citep{1980ApJS...43..469T}. The function
$\varepsilon(\Dnu)$ enables the identification of the radial modes
\begin{equation}\label{ridge}
\nu_{n,\ell=0} = [n+ \varepsilon(\Dnu)] \;\Dnu,
\end{equation}
with $\Dnu$ the large separation averaged in the frequency range
$[\numax-\dnuenv, \numax+\dnuenv]$. For simplicity, we introduce
the \emph{reduced frequency}, dimensionless and corrected for the
$\varepsilon$ term of the Tassoul equation:
\begin{equation}\label{reduit}
n' = \nu / \Dnu - \varepsilon (\Dnu) .
\end{equation}
According to \cite{2010ApJ...723.1607H} and
\cite{2011A&A...525L...9M}, $\varepsilon$ is mainly a function of
the large separation. For radial modes, the reduced frequency is
very close to the radial order, except for possible very small
secondary-order terms in $\varepsilon(\Dnu)$ that do not depend
directly on $\Dnu$ and do not hamper the mode identification.

The determination of the evolutionary status of the giants
(Sect.~\ref{statut}), namely the measurement of the g-mode spacing
of $\ell=1$ mixed modes, is then achieved with the automated
method described by \cite{2011A&A...532A..86M}. This could be done
for about 674 targets out of 1043 with high signal-to-noise ratio
time series. This method measures the g-mode spacing of the mixed
modes from the Fourier spectrum of the oscillation spectrum
windowed with narrow filters centered on the expected locations of
each pure $\ell=1$ pressure mode.

\end{appendix}
\begin{acknowledgements}
Funding for this Discovery mission is provided by NASA's Science
Mission Directorate. YE, SH and WJC acknowledge financial support
from the UK Science and Technology Facilities Council. SH
acknowledges financial support from the Netherlands Organisation
for Scientific Research (NWO). NCAR is supported by the National
Science Foundation. DS and TRB acknowledge support by the
Australian Research Council. TK acknowledges the support of the
FWO-Flanders under project O6260 - G.0728.11.

\end{acknowledgements}

\bibliographystyle{aa} 
\bibliography{biblio_energy}

\end{document}